\documentstyle[elsart12,osa,epsf,eqsecnum]{revtex}
\begin{document}


\voffset1.5cm
\title{Eikonal Evolution and Gluon Radiation}
\author{Alex Kovner$^{1,2}$ and Urs Achim Wiedemann$^1$}

\address{$^1$Theory Division, CERN, CH-1211 Geneva 23, Switzerland\\
and\\
$^2$Department of Mathematics and Statistics, University of Plymouth,
Plymouth PL4 8AA, UK}

\date{\today}
\maketitle

\begin{abstract}
We give a simple quantum mechanical formulation of the eikonal 
propagation approximation, which has been heavily used in recent years
in problems involving hadronic interactions at high energy. This 
provides a unified framework for several approaches existing in the 
literature. We illustrate this scheme by calculating the total, 
elastic, inelastic and diffractive DIS cross sections, as well as
gluon production in high energy hadronic collisions. From the
$q\bar{q}g$-component of the DIS cross sections, we straightforwardly
derive low x 
evolution equations for inelastic and diffractive DIS distribution 
functions. In all calculations, we provide all order 
$1/N$ corrections to the 
results existing in the literature.
\end{abstract}
\vspace{1cm}
Preprint number: CERN-TH/2001-162


\section{Introduction}
\label{sec1}

In recent years a lot of work has been done aimed at understanding 
the physics of hadronic interactions in a dense (nuclear) environment. The 
questions addressed in this context range form DIS at low $x$ to high 
energy scattering on nuclear targets to heavy ion collisions. Experimental 
motivations come from the low $x$ data measured at the DESY collider
and the Fermilab Tevatron and, of course, heavy ion collision experiments 
at RHIC and LHC. An additional theoretical motivation is that the dense 
partonic environment provides a link between straight perturbative QCD 
of factorization theorems and Dokshitzer-Gribov-Lipatov-Altarelli-Parisi 
(DGLAP) evolution, and the still poorly understood soft QCD 
physics. The main idea is that high partonic density might provide 
an effective infrared cutoff, so that the genuine infrared strongly 
interacting region becomes relatively unimportant. The 
contribution of the infrared region to various observables 
which are not infrared safe in the usual perturbative sense, then may be 
suppressed by powers of $\Lambda/\mu$, where $\Lambda$ is the infrared scale
and $\mu$ is the scale roughly measuring the strength of the gluon fields 
in the dense partonic system, $\mu\propto gA$. If $\mu$ is large enough, 
the coupling constant governing important interactions is small, and
many non infrared safe quantities may become in principle calculable. 
Although we are still far from the quantitative understanding of this 
interesting regime, progress is being continuosly made in different
approaches~\cite{glr,bdmps-glv,buchmuller,mv,al-yuri,levin,bk,jklw,kaidalov,makhlin,frankfurt,kv99,lm2000,krt01,np01}. 

In this paper our aim is to give a simple physical description and 
provide a transparent calculational framework to what one may call 
eikonal propagation approach. The eikonal picture views a high energy 
hadronic reaction as a projectile, consisting of a (usually small) 
number of partons impinging on a large target. The target could be 
either a nucleus in its rest frame, or a high energy hadron which has 
already developed many partons in its wave function. The interaction 
between the projectile partons and the target fields is assumed to be 
eikonal: that is the projectile partons propagate through the target 
without changing their transverse position but picking up an eikonal phase. 
As a result of the phase differences accumulated by the partons, the initial 
wave function decoheres and a different hadronic final state is produced.
The target in this picture is described by an ensemble of gluon field
configurations, and the averaging over this ensemble determines the cross
section for various processes. 

Eikonal physics underlies recent approaches towards calculating the 
produced particle multiplicities in high energy hadron-nucleus~\cite{al-yuri} 
and nucleus-nucleus~\cite{K00} collisions. It is applied in the 
calculations of radiative jet energy loss in hot and cold nuclear 
matter~\cite{bdmps-glv} as well as in many studies of low $x$
DIS. A particularly lucid discussion with many explicit calculations 
is given by Buchm\"uller et al.~\cite{buchmuller}. Moreover, as 
discussed in Ref.~\cite{guilherme}, eikonal propagation is closely 
linked to the low $x$ evolution of Refs.~\cite{bk,jklw}. 
On the other hand, the eikonal approach has been critisized in 
\cite{frankfurt}, since it does not take properly into account 
production of diffractive intermediate states, which should 
be equally important at high energy~\cite{gribov}.
One of the main motivations for our work is indeed to set up a simple
framework of eikonal propagation which can be extended
to include diffractive production. While the present paper
does not include diffractive production, we view it as a 
step towards this goal~\cite{inprep}. 

This paper is organized as follows:
In Section 2 we give an explicit quantum mechanical description of 
the eikonal evolution and show how to calculate various cross sections 
from the knowledge of the outgoing wave function of the projectile 
that has undergone eikonal scattering. 
We explain how to perform averages over the gluon fields of the target.
It turns out that the correct averaging procedure amounts to averaging on 
the amplitude level for the elastic and inelastic cross section, but on 
the cross section level for the diffractive one.

As a first illustration of our calculational scheme, we derive in 
Section 3 the gluon radiation in a hadron-nucleus scattering. This 
calculation reproduces the results of Ref.~\cite{al-yuri} for $q-A$ 
and $q\bar q -A$ systems.

In Section 4, we discuss Deeply Inelastic Scattering on a large target.
Our calculations are consistent with the results of Buchm\"uller et 
al.\cite{buchmuller}, and provide the $1/N$ corrections to them. 
The most interesting quantity here is the diffractive cross section.
We explain why physically the $1/N$ corrections in this case 
are precisely what distinguishes between the elastic cross 
section calculated in \cite{buchmuller,yuri-larry} and
the diffractive, or large rapidity gap cross section that we calculate here.
We calculate the diffractive cross section due to the $\bar q q$ as well as
to the $\bar q q g$
component of the virtual photon.

In Section 5 we consider the low $x$ evolution in DIS. For the total 
photoabsorption cross section our results immediately yield the operator 
form of the low $x$ evolution equation of Ref.~\cite{bk}.
We derive $1/N$ corrections to Kovchegov's version of this equation and
show that these corrections express the fact that the fundamental dipoles
do not propagate through the target independently beyond the leading 
order in $1/N$. We also derive an analogous evolution equation for the 
diffractive cross section generalizing the results of Ref.~\cite{yuri-genya}.
Here we again obtain the operator form of the evolution, which is a new 
result. We show that in the large $N$ limit this equation reduces to that 
of Ref.~\cite{yuri-genya} and we derive all-order $1/N$ corrections to 
this equation within our model for the averaging over the target fields.
We point out that both the leading and all-order $1/N$ evolution is not
an independent equation. Once the evolution of the total cross 
section is determined, the diffractive cross section is obtained
by direct integration of the known function of rapidity.

Finally in Section 6 we summarize our results.

\section{Quantum mechanics of eikonal propagation}
\label{sec2}

Consider an energetic hadronic projectile impinging on a large nuclear 
target. The projectile is characterised by a wave function, in which 
the relevant degrees of freedom are the transverse positions and colour 
states of the partons,
\begin{equation}
  \Psi_{in} = \sum_{\{\alpha_i,x_i\}}\, \psi(\{\alpha_i, x_i\})\, 
  \vert\{\alpha_i,x_i\}\rangle\, .
  \label{2.1}
\end{equation}
The colour index $\alpha_i$ can belong to the fundamental,
antifundamental or adjoint representation of the colour $SU(N)$ group, 
corresponding to quark, antiquark or gluon in the wavefunction. 
In what follows we will consider wave functions with a small number 
of partons.

\subsection{The projectile biased observables} 
\label{sec2a}

At high energy the propagation time through the target is short,
and thus partons propagate independently of each other. For the same 
reason the transverse positions of the partons do not change during 
the propagation. The only effect of the propagation is that the wave 
function of each parton acquires an eikonal phase due to the interaction 
with the glue field of the target. Thus the projectile emerges form the 
interaction region with the wave function
\begin{equation}
  \Psi_{out}={\cal S}\Psi_{in}=
  \sum_{\{\alpha_i,x_i\}}\psi(\{\alpha_i, x_i\})
  \prod_iW(x_i)_{\alpha_i \beta_i}\, \vert\{\beta_i,x_i\}\rangle\, .
  \label{2.2}
\end{equation}
Here ${\cal S}$ is the $S$-matrix, and the $W$'s are Wilson lines
\begin{equation}
  W(x_i)={\cal P}\exp\{i\int dz_-T^aA^+_a(x_i, z_-)\}
  \label{2.3}
\end{equation}
with $A^+$ - the gauge field in the target and $T^a$ - the generator of 
$SU(N)$ in a representation corresponding to a given parton. Thus the 
relative phases between components of the wave function change, and the 
state that emerges after the target is no longer an eigenstate of the 
strong interaction Hamiltonian (as the incoming state is assumed to be)
but rather a superposition thereof. 

Given this explicit law for the evolution of the projectile
wave function, one can 
calculate various observables related to the scattering process.
For example, up to the factor of total flux, the inelastic cross 
section is proportional to the probability that the outgoing state 
is different from the incoming one.
This probability is given by the norm 
\begin{equation}
  P^{\rm proj}_{\rm inel}=|\delta\Psi|^2\, ,
  \label{2.4}
\end{equation}
where
\begin{eqnarray}
  |\delta\Psi\rangle &=& 
\left[1-|\Psi_{in}\rangle\langle\Psi_{in}|\right]|\Psi_{out}\rangle\nonumber\\
&=&|\Psi_{out}>-s|\Psi_{in}\rangle\, ,
  \label{2.5}
\end{eqnarray}
and $s$ is the $S$-matrix element, or overlap
\begin{equation}
  s=\langle\Psi_{in}|\Psi_{out}\rangle\, .
  \label{2.6}
\end{equation}
Thus
\begin{equation}
  P^{\rm proj}_{\rm inel}=1-\vert\langle\Psi_{in}|\Psi_{out}\rangle\vert^2
  \label{2.7}
\end{equation}
The probability $P^{proj}_{inel}$ for a given configuration of the 
target fields of course depends on the impact parameter, that is the 
position of the projectile in the transverse plain. To get the cross 
section one simply integrates the probability over the impact parameter
\begin{equation}
  \sigma^{\rm proj}_{\rm inel}=\int d^2b P^{\rm proj}_{\rm inel}(b)\, .
  \label{2.8}
\end{equation}
Note that the probability $P^{\rm proj}_{\rm inel}$ as defined above
does not include processes for which 
the projectile remains intact but the target is excited or breaks into 
fragments that stay close to the original rapidity. By its very definition,
$P^{proj}_{inel}$ is the probability that the {\it projectile} scatters 
inelastically. We refer to it as a {\it projectile biased observable}
and we denote it by a superscript ``proj''. The formulation which
removes this bias will be given in section~\ref{sec2c}.

More generally, the incoming state may have internal degrees of freedom
which we collectively denote by $\gamma$. This can be spin or 
(perturbatively) global colour. In that case the 
inelastic probability is the probability that the outcoming state is
perpendicular to any one of the states  $\Psi_{in}(\gamma)$. Technically
this means that $\delta\Psi$ is defined by subtracting form $\Psi_{out}$
its projection onto a subspace spanned by $\gamma$
\begin{equation}
  \vert\delta\Psi\rangle =
  \left[1-\sum_\gamma|\Psi_{in}(\gamma)\rangle\, 
  \langle\Psi_{in}(\gamma)|\right]|\Psi_{out}\rangle\, .
\label{2.9}
\end{equation}
For orthonormal states $\Psi_{in}(\gamma)$ the operator
\begin{equation}
  p_\gamma=1-\sum_\gamma|\Psi_{in}(\gamma)\rangle\,
  \langle\Psi_{in}(\gamma)|
  \label{2.10}
\end{equation}
is the projector on the subspace orthogonal to the possible initial
states of the projectile hadron. 

One can also calculate other, less inclusive observables. For example 
the number of gluons with momentum $k$ produced in the scattering event 
is given by
\begin{equation}
  N_{\rm prod}({\bf k}) =
  \langle\delta\Psi| a^\dagger(k)\, a(k)|\delta\Psi\rangle\, .
  \label{2.11}
\end{equation}
In principle, the calculation of any observable in an inelastic
process has to allow
for the evolution of the state after the interaction to the time when the 
measurement is performed, that is $t=\infty$. Thus the relevant observable
should be calculated not in the state $|\delta\Psi>$, but rather in
$U(0,\infty)|\delta\Psi>$, where $U(0,\infty)$ is the free evolution 
operator from the time zero (immediatelly after the interaction)
to infinity. Being a unitary operator, it 
cancels against its conjugate in the calculation of the cross section 
Eq.~(\ref{2.4}), but it affects some less inclusive observables, such as 
the number of produced gluons (\ref{2.11}). In the quantities we calculate 
in this paper its contribution is always a perturbative correction of 
higher order than the accuracy to which we are working, and we do not 
consider this type of correction. As will become apparent from our 
calculations in the later sections, this does not mean that we neglect 
contributions which in the Feynman diagram language correspond to 
``emission after interaction''. Those in the present framework arise from the 
``overlap'' contributions, proportional to the overlap $s$ of Eqs.~
(\ref{2.5},\ref{2.6}).

In contrast to the inelastic cross section, the total cross section
includes the probability for an elastic process, for which the 
outcoming state is the same as the incoming one up to a phase 
shift. In this case, one includes in the calculation all components
of the outgoing wave function which differ from the incoming one,
\begin{equation}
  \vert \delta\Psi_{\rm tot}\rangle=\vert\Psi_{\rm out}\rangle
  -\vert\Psi_{\rm in}\rangle\, ,
  \label{2.12}
\end{equation}
and
\begin{eqnarray}
  \sigma_{\rm tot}&=&
  \int d^2b\, \langle\Psi_{\rm out}(b)-\Psi_{\rm in}(b)\vert
  \Psi_{\rm out}(b)-\Psi_{\rm in}(b)\rangle\nonumber \\
  &=&
  \int d^2b[2-\langle\Psi_{\rm out}(b)\rangle|\Psi_{\rm in}(b)\rangle-
  \langle\Psi_{\rm in}(b)\rangle|\Psi_{\rm out}(b)\rangle]\, .
  \label{2.13}
\end{eqnarray}

The elastic cross section measures the probability of a process in which 
the product of the collision is in the same state as the initial one but
with a nontrivial phase shift. This is defined through
\begin{equation}
  \vert\delta\Psi_{\rm el}\rangle
  =[\langle\Psi_{\rm in}\vert\Psi_{\rm out}>-1]\vert\Psi_{\rm in}\rangle\, ,
  \label{2.14}
\end{equation}
and
\begin{equation}
  \sigma_{\rm el}^{\rm proj}=
  \int d^2b\, 
  \langle\delta\Psi_{\rm el}\vert\delta\Psi_{\rm el}\rangle=
  \int d^2b\, \vert 
  1-\langle\Psi_{\rm out}(b)\rangle\vert\Psi_{\rm in}(b)\rangle|^2\, .
  \label{2.15}
\end{equation}

Another interesting quantity
is the diffractive cross section. This includes all elastic 
and inelastic processes for which the outgoing state is different from 
the incoming one and is in a colour singlet,
\begin{equation}
  \vert\delta\Psi_{\rm diff}\rangle={\cal P}_{\rm singlet}\vert\Psi_{\rm out}
\rangle-\vert\Psi_{\rm in}\rangle
  \label{2.16}
\end{equation}
and
\begin{equation}
  \sigma_{\rm diff}=
  \int d^2b\, \langle{\cal P}_{\rm singlet}\Psi_{out}(b)-\Psi_{in}(b)|
  {\cal P}_{\rm singlet}\Psi_{out}(b)-\Psi_{in}(b)\rangle\, .
  \label{2.17}
\end{equation}
The operator ${\cal P}$ is the projection operator 
on the colour singlet states.

\subsection{The target averages}
\label{sec2b}
The only property of the gluon field in the 
target assumed in the
preceding discussion, is  that it
does not vary during the time of the interaction. This is 
justified since the time of the propagation of an energetic projectile 
through the target is shorter than the natural time scale on which 
the target fields vary. The target is, however, characterised by a 
wave function which has a spread in field space. The relative phases 
of different components of this wave function are not relevant in the 
eikonal approximation, but the probability distribution for different 
field strengths is relevant indeed. 
In other words, while each scattering event 
happens on a frozen field profile, in order to calculate the cross 
section one has to average over the field ensemble which represents 
the wave function of the target.

In this paper we adopt the same ensemble averaging procedure as 
in Refs.~\cite{bdmps-glv,nucl,al-yuri}. The target is pictured as
an assembly of sources of the colour field extended in both the 
longitudinal and transverse directions. The sources are independent 
of each other and thus completely decorrelated. The longitudinal and 
transverse coordinates of the sources are randomly distributed within 
the target's longitudinal and transverse extent. The field emitted by 
each source is small, and to leading order in the field strength,
the projectile can exchange only one or two gluons with each one of 
the sources. Pictorially, the average of, say a product of two Wilson loops
can be represented as in Fig.1.

\begin{figure}[h]\epsfxsize=8.7cm 
\centerline{\epsfbox{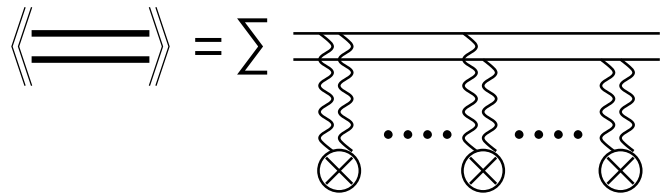}}
\vspace{0.5cm}
\caption{The target average of the product of two Wilson lines, 
denoted by thick lines.
The fields $A^+$ in $W$'s are distributed with Gaussian weights
local in the longitudinal coordinate according to (\ref{2.18}).}
\label{fig2}
\end{figure}

Put into formulae, this means that the target average
of the two-point function of the vector potential $A^+$,
which enters  all our calculations through Eqs.~(\ref{2.2},\ref{2.3})
is
\begin{equation}
  \langle\langle A^{+a}({\bf x}, x^-) A^{+b}({\bf y}, y^-)
  \rangle\rangle
  =\delta^{ab}\, \delta(x^--y^-)\, B({\bf x,y},x^-)\, .
  \label{2.18}
\end{equation}
All final results will depend on one particular linear combination
of these averages
\begin{equation}
  v({\bf x,y})=\int dx^-{1\over 2}[
  B({\bf x,x},x^-)+ B({\bf y,y},x^-) - 2B({\bf x,y},x^-)]\, .
  \label{2.19}
\end{equation}
For simplicity of notation we will sometimes consider the field ensemble to be
translationally invariant in the transverse plain within some radius $R$, 
and will take all fields to vanish outside this radius. In this 
approximation, $v$ depends only on the difference of its coordinates.

With this averaging procedure, the target averages can be explicitly 
calculated. This is done for a variety of observables in Appendix C. 
The simplest averages we will need are those corresponding to 
``fundamental dipole'' and ``adjoint dipole'' scattering amplitudes
\begin{eqnarray}
\frac{1}{N}
\langle\langle{\rm Tr}\left[ W^{F\dagger}({\bf x})
                   W^{F}({\bf y})\right]\rangle\rangle
  & =& \exp\left[-C_F\, v({\bf x}-{\bf y})\right]\, ,
   \label{2.20}\\
    \frac{1}{N^2 - 1}\, 
  \langle\langle
  {\rm Tr}\left[ W^{A\, \dagger}({\bf y})\, W^A({\bf x}) \right]
  \rangle\rangle
  &=& \exp\left[-C_A\, v({\bf x}-{\bf y})\right]\, ,
\label{2.21} 
\end{eqnarray}
with the Wilson loops in the fundamental and adjoint representations 
respectively.

In most of the paper we will not specify the form of $B$, but sometimes it is
useful to keep in mind the following simple form \cite{al-yuri,nucl}
\begin{equation}
  B({\bf x,y})={\bf x}\cdot{\bf y}\, \mu^2(x^-)\, ,
  \label{2.22}
\end{equation}
where $\mu$ is a slowly varying function of $x^-$. The independence of $\mu$ 
on ${\bf x}$ inside the radius of the target corresponds to the leading 
logarithmic approximation in the following sense. We are working in the 
gauge where the only nonvanishing component of the vector potential
in the target is $A^+$. In this gauge the vector potential is related
to the colour electric field by the following relation which is local in 
$x^-$:
\begin{equation}
  A^+({\bf x})=\int^{\bf x}_{\bf x_0} dx_iE_i\, .
  \label{2.23}
\end{equation}
The leading logarithmic approximation assumes that the scale of the 
variation of the color electric field is much larger than the transverse 
distances at which the field is probed (transverse momentum ordering). 
In this approximation, we have
\begin{equation}
  A^+({\bf x})=({\bf x-x_{0}})\cdot{\bf E}
  \label{2.24}
\end{equation}
and thus choosing ${\bf x_0}=0$, (\ref{2.18}) with (\ref{2.22}) follow
for isotropic distribution of $E_i$. [It is in fact not 
essential to choose ${\bf x_0}=0$, since the dependence on the base
point ${\bf x_0}$ will drop out from all physical quantities. All
physical observables, as we shall see
depend only on the difference $A^+({\bf x})-A^+({\bf y})$
of the vector potentials.]
In this leading logarithmic approximation, the dipole scattering amplitude
of eq.(\ref{2.20}) is
\begin{equation}
\langle\langle\frac{1}{N}
    {\rm Tr}\left[ W^{F\dagger}({\bf x})
                   W^{F}({\bf y})\right]\rangle\rangle
  = \exp\left[-{({\bf x}-{\bf y})^2\over 8}Q^2_s\right]\, ,
  \label{2.25}
\end{equation}
with the saturation momentum
\begin{equation}
  Q^2_s=4C_F\int dx^-\mu^2(x^-)\, .
  \label{2.26}
\end{equation}

Thus to reiterate, the procedure of calculating any observable in 
the eikonal approximation is the following. For a given initial state,
we calculate its eikonal propagation according to (\ref{2.2}). Next 
we ``project out'' the subspace which is spanned by the possible initial 
states, according to (\ref{2.9}). Then we calculate the observable 
as in (\ref{2.11}). Then we perform the averaging over the initial state if 
this is needed - for example the averaging over the initial polarization in 
the case of unpolarized scattering. And finally we average over the target 
field ensemble in the two gluon exchange approximation with the only 
nontrivial target average specified in (\ref{2.18}) (and (\ref{2.22})).
%
\subsection{With due respect to the target}
\label{sec2c}

In the eikonal scheme described above, one does not follow the details 
of the dynamics on the target side, since the target is modelled by a
field distribution rather than being assigned a wave function of its own. 
As a result, the inelastic cross section (\ref{2.8}) calculated
in this way is underestimated since it accounts only for processes
in which the {\it projectile} scatters inelastically. Other inelastic
processes in the usual sense, in which e.g. the projectile remains intact 
but the target breaks up, are not included in $\sigma_{inel}^{proj}$.

\begin{figure}[h]\epsfxsize=8.7cm 
\centerline{\epsfbox{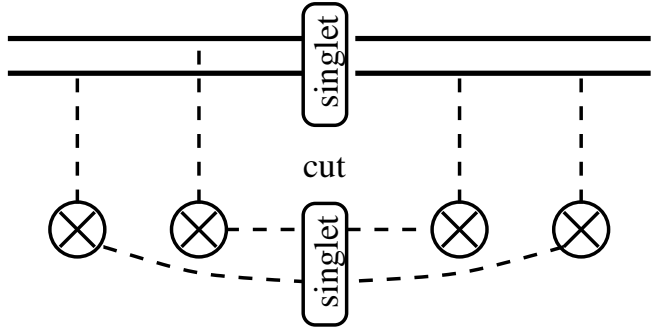}}
\vspace{0.5cm}
\caption{Schematic picture of a diffractive contribution to the
$q\bar{q}$ scattering cross section which does contribute to
the inelastic scattering cross section.
}\label{fig1}
\end{figure}

In the same way, the cross section $\sigma_{el}^{proj}$ in (\ref{2.15})
accounts for the probability that the {\it projectile} undergoes elastic 
scattering. Thus its definition includes the probability of
inelastic processes where only the target is excited.
Such ``diffractive'' processes are correctly included in the
diffractive cross section $\sigma_{diff}$ in (\ref{2.17}), but in 
the above, projectile biased discussion, they are shifted from their 
rightful place in $\sigma_{inel}$ into $\sigma_{el}$.
As we shall see in our explicit calculations, these contributions
are suppressed by powers of $1/N$ in the situation where the projectile
is much smaller than the target. It is however unclear to us whether the 
same is true when projectile and target are roughly of the same size. 

In order to have the proper separation between elastic and inelastic 
contributions, we consider now the role of the target wave function. 
We observe first that the target field averaging described above 
amounts to calculating averages of the appropriate Wilson loops in the 
target wave function. Thus for example
\begin{eqnarray}
    \langle\langle
    {\rm Tr}\left[ W^{F\dagger}({\bf x})
                   W^{F}({\bf y})\right]\rangle\rangle&\equiv&
    \langle\Psi^t_{in}|
    {\rm Tr}\left[ W^{F\dagger}({\bf x})
                   W^{F}({\bf y})\right]|\Psi^t_{in}\rangle\nonumber \\
    &=& \int DA{\rm Tr}\left[ 
    W^{F\dagger}({\bf x})W^{F}({\bf y})\right]|\psi^t[A]|^2\, ,
    \label{2.27}
\end{eqnarray}
where $\Psi^t_{in}$ is the initial state target wave function
which in the gluon field basis has the form 
\begin{equation}
  |\Psi^t_{in}\rangle=\psi^t[A]|A\rangle\, .
  \label{2.28}
\end{equation}
We can now parallel the discussion of subsection~\ref{sec2a} 
paying proper attention to the target wave function effects.
The initial wave function of the projectile+target system is
\begin{equation}
  \Psi_{in} = \sum_{\{\alpha_i,x_i\}}\, \psi(\{\alpha_i, x_i\})\, 
  \psi^t[A]\vert\{\alpha_i,x_i\}\rangle\, \otimes
  \vert A\rangle\, .
\label{2.29}
\end{equation}
The outgoing wave function is
\begin{equation}
  \Psi_{out}={\cal S}\Psi_{in}=
  \sum_{\{\alpha_i,x_i\}}\psi(\{\alpha_i, x_i\})
  \prod_iW(x_i)_{\alpha_i \beta_i}\, \psi^t[A]\vert\{\beta_i,x_i\}\rangle\,
  \otimes
  \vert A\rangle\,  .
  \label{2.30}
\end{equation}
Although the incoming wave function is a product of the wave functions
of the target and the projectile, the outgoing wave function is not 
factorizable, since the eikonal factors $W$ depend both, on the 
coordinates of the projectile particles and on the target fields.
The $S$-matrix element, or overlap that determines the inelastic
scattering probability (\ref{2.4}) reads now
\begin{equation}
  s=\langle\Psi_{in}|\Psi_{out}\rangle\,
  =\langle[\langle W\rangle_t]\rangle_p\, ,
  \label{2.31}
\end{equation}
where the incoming and outgoing wave functions (\ref{2.29}) and (\ref{2.30}) 
contain both the projectile and the target parts, rather than just 
the projectile parts as in (\ref{2.1}) and (\ref{2.2}).
Here we denoted the Wilson loop factors arising in the $S$-matrix
collectively by $W$, $\langle\rangle_t$ denotes the averaging 
over the target fields (wave function) as in (\ref{2.27}), 
and $\langle\rangle_p$ is the expectation value 
with respect to the incoming projectile wave function\footnote{Here,
in order to project out the projectile states with $p_\gamma$, we
proceed in exactly the same way as explained in eqs.(\ref{2.9},\ref{2.10}). 
We thus assume that the only degrees of freedom that characterise the target
are the vector potentials, and that there is therefore no need to 
perform an analogous projection on the target wave function.}. 
The same goes for the elastic probability of eq.(\ref{2.14}) and the total
probability eq.(\ref{2.12}).
Thus we find
\begin{eqnarray}
  P_{inel}&=&1-|s|^2\, , 
  \label{2.32}\\
  P_{el}&=&|1-s|^2\, ,
  \label{2.33}\\
  P_{tot}&=&P_{inel}+P_{el}=2-s-s^*\, ,
  \label{2.34}\\
  \sigma_{inel,el,tot}&=&\int d^2b P_{inel,el,tot}(b)\, .
  \label{2.35}
\end{eqnarray}
The $S$-matrix element $s$, given by (\ref{2.31}), already contains 
the target average.

We therefore see that in order to calculate the true elastic and
inelastic cross sections, the ``target field averaging'' has to be 
performed on the amplitude level rather than on the
cross section level as is sometimes thought\cite{yuri-larry}.
The difference between these averaging procedures 
has a clear physical meaning: they differ in the treatment of the
diffractive states close to the rapidity of the target (see Fig.~\ref{fig1}).

Not all quantities are affected by this more careful treatment 
of the target wave 
function in calculating the overlap (\ref{2.31}).
The total cross section is not sensitive to this, 
since it is linear in $s$ and therefore contains in any case only 
one averaging over the target fields. Physically this is because 
it does not care about whether we assign the target side diffractive 
contribution to the elastic or the inelastic probability. The 
diffractive cross section is not affected since we defined diffractive 
processes as those in which the projectile wave function emerges 
from the interaction region in a colour singlet state. This definition
does not refer to the final state of the target, and thus the target 
averaging should only be performed once, on the cross section level.
The number of produced gluons is also unaffected if we are interested 
in gluons at rapidities far from those of the target.

In the rest of this paper, we apply the eikonal approximation to a 
variety of simple cases. Whenever we are interested in the cross
sections, we will calculate them according to both definitions: 
the one referring only to the projectile as described in 
subsection~\ref{sec2a} and the one including the proper treatment
of the target wave function, as described above. The projectile biased 
quantities we will denote by the superscript ``proj''. We calculate both 
quantities in order to illustrate the differences between them. We 
stress, however, that the physical inclusive cross sections are given 
by the expressions eq.(\ref{2.35}). In the eikonal approximation,
the proper calculation of the elastic and inelastic cross section involves
the target field averages on the amplitude level.

\section{Gluon production}
\label{sec3}
As a first illustration of our formalism, we present
a toy calculation of the number of gluons produced
in the high energy scattering of a quark off a nuclear target. 
%
\subsection{Gluon production in q+A}
\label{sec3a}
The incoming state in this case is a single quark idealized as the 
Fock state $ \vert \alpha\rangle$ of the bare quark, supplemented 
by the coherent state of quasireal gluons which build up the 
Weizs\"acker-Williams field $f({\bf x})$. This can be written as 
a coherent state of a gluon field - the result of evolving the
bare quark state by free evolution from time $-\infty$ to the moment of 
the interaction
\begin{eqnarray}
  U_-\, \vert \alpha\rangle &=& 
  \exp\left(- \int d{\bf x}\, \vert f({\bf x})\vert^2 
   + i \int d{\bf x}\,d\xi f({\bf x})\, 
    [a_d({\bf x},\xi) + a_d^\dagger({\bf x},\xi)]\, T^d \right)\, 
   \vert \alpha\rangle\nonumber \\        
  &\approx&      
  \vert \alpha\rangle + \int d{\bf x}\,d\xi f({\bf x})\, 
  T^b_{\alpha\, \beta}\, \vert \beta\, ; b({\bf x},\xi)\rangle\, .
  \label{3.1}
\end{eqnarray}
Here, $U_-$ denotes the unitary free time evolution from $-\infty$ 
up to the target, $a_d({\bf x},\xi)$, $a_d^\dagger({\bf x},\xi)$ are the 
annihilation and creation operators for gluons of colour $d$ at
transverse positon ${\bf x}$, and rapidity $\xi$; 
and $T^d$ are generators of the
fundamental representation of $SU(N)$. Lorentz and spin indices are 
suppressed. We work in the light cone gauge appropriate to the projectile, 
that is $A^-=0$. In this gauge the gluon field of the projectile is 
the Weizs\"acker-Williams field
\begin{equation}
  A^i({\bf x})\propto  \theta(x^-)\, f_i({\bf x})\, ,\qquad \qquad
  f_i(x_\perp)\propto g{{\bf x_i}\over {\bf x}^2}\, ,
  \label{3.2}
\end{equation}
where $x^-=0$ is the light cone coordinate of the quark in the wave 
function. The integration over the rapidity of the gluon in the wave 
function (\ref{3.1}) goes over the gluon rapidities smaller than that 
of the quark. In the following we will suppress the rapidity label. 
We are interested in calculating the distribution of produced gluons 
per unit rapidity and since the wave function does not depend on 
rapidity, this distribution is obviously flat.
 
The Weizs\"acker-Williams field is proportional to the QCD coupling
and is considered to be small. The higher powers of
$\vert f({\bf x})\vert^2$ are thus negligible and the 
approximation (\ref{3.1}) applies.

The interaction $S$ of the projectile wavefunction (\ref{3.1})
with the target results in the phase shifts described by eikonal
Wilson lines,
\begin{eqnarray}
 \Psi_{out} &=& S\, U_-\,  \vert \alpha\rangle 
 \nonumber \\
 &=& 
  W^F_{\alpha\, \gamma}({\bf 0})\, \vert\gamma\rangle +
  \int   d{\bf x}\, f({\bf x})\, T_{\alpha\, \beta}^b
  W_{\beta\, \gamma}^F({\bf 0})\, W_{b\, c}^A({\bf x})\, 
  \vert \gamma\, ;c({\bf x})\rangle\, .
  \label{3.3}
\end{eqnarray}
Here $W^F({\bf 0})$ and $W^A({\bf x})$ are the Wilson lines in the 
fundamental and adjoint representations respectively, corresponding 
to the propagating quark at the transverse position ${\bf x}_q={\bf 0}$ 
and gluon at ${\bf x}_g={\bf x}$.

Gluon production being an inelastic process, we have to project out of 
the outgoing wave function the component lying in the subspace spanned 
by the incoming states with arbitrary colour orientation $\alpha$. Thus 
the index $\gamma$ in the projection operator (\ref{2.10}) has here the 
meaning of the quark colour index,
\begin{eqnarray}
  \vert \delta \Psi_\alpha \rangle =
  U_+\,\left [\vert \Psi_{out}\rangle -
  \sum_\gamma   U_-\, \vert \gamma\rangle \langle \gamma 
  U_-^\dagger\, U_+^\dagger\, \vert \Psi_{out}\rangle\, \right]\, .
  \label{3.4}
\end{eqnarray}
To lowest order in $f^2$ only the no-gluon component of $\Psi_{out}$ 
contributes in the last (overlap) term,
\begin{eqnarray}
 &&\vert \delta \Psi_\alpha \rangle =  U_+\,
  \int   d{\bf x}\, f({\bf x})\,\left [ T_{\alpha\, \beta}^b
  W_{\beta\, \gamma}^F({\bf 0})\, W_{b\, c}^A({\bf x})\, -
 T_{\beta\, \gamma}^c
  W_{\alpha\, \beta}^F({\bf 0})\,\,  \right ]
  \vert \gamma\, ;c({\bf x})\rangle\,
\label{3.5}\\
&&
\epsfxsize=12.0cm 
\centerline{\epsfbox{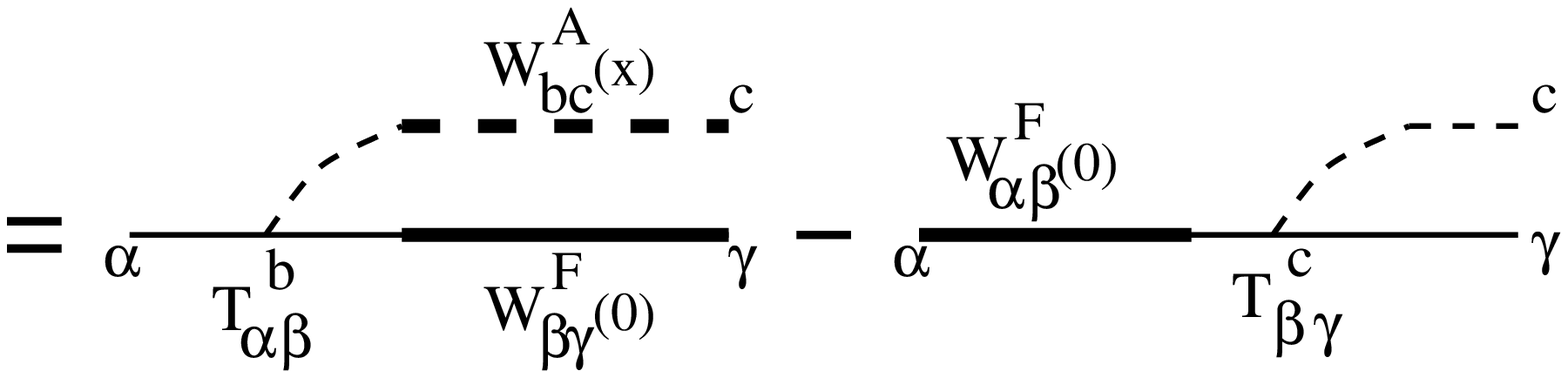}}
\vspace{-1cm}
\nonumber 
\end{eqnarray}
As illustrated by the diagrammatic representation of this equation,
the structure of the second term corresponds precisely to 
``emission after interaction''. By this we mean 
that the gluon is emitted from the initial quark state rotated by
the eikonal factor $W^F$, while there is no eikonal factor $W^A$ that 
accompanies the gluon propagating through the target.
Thus this term is the exact counterpart of the
Feynman diagrams containing gluon emission from final state quark.
Another point to note is that since the r.h.s. of (\ref{3.5})
has an explicit factor $f$, to lowest order in $f^2$ the evolution
operator $U_+$ in (\ref{3.5}) 
has to be approximated by unity [{\it cf} eq.(\ref{3.1})],
as noted in the previous section.

The number spectrum of produced gluons is obtained from (\ref{3.5})
by calculating the expectation value of the number operator in the 
state $\delta \Psi_\alpha$, averaged over the incoming colour index 
$\alpha$. After some colour algebra (see appendix~\ref{appb} for 
technical details) we obtain
\begin{eqnarray}
  && N_{\rm prod}({\bf k}) = \frac{1}{N} \sum_\alpha
  \langle \delta \Psi_\alpha\vert a_d^\dagger({\bf k})\, 
  a_d({\bf k})\vert\, \delta\Psi_\alpha \rangle
  \nonumber\\
  && \quad 
  = C_F\, \int d{\bf x}\, d{\bf y}\, e^{i{\bf k}\cdot({\bf x}-{\bf y})}
  f({\bf x})\, f({\bf y})
  \Bigg[ 1 - \frac{1}{N^2 - 1}\, 
  \langle\langle
  {\rm Tr}\left[ W^{A\, \dagger}({\bf x})\, W^A({\bf 0}) \right]
  \rangle\rangle_t
  \nonumber\\
&& \qquad \qquad \qquad \qquad \qquad \qquad  - \frac{1}{N^2 - 1}\, 
  \langle\langle
  {\rm Tr}\left[ W^{A\, \dagger}({\bf y})\, W^A({\bf 0}) \right]
  \rangle\rangle_t
  \nonumber \\
&& \qquad \qquad \qquad \qquad \qquad \qquad
  +\frac{1}{N^2 - 1}\, 
  \langle\langle
  {\rm Tr}\left[ W^{A\, \dagger}({\bf y})\, W^A({\bf x}) \right]
  \rangle\rangle_t\Bigg]\, ,
  \label{3.7}
\end{eqnarray}
Here, all the information about the scattering properties of the target
are encoded in the target average of the product of adjoint Wilson loops
which are determined by (\ref{2.18}) and (\ref{2.22}). The explicit 
calculation of this as well as some more complicated Wilson loop averages 
is given in Appendix C.
The result is
\begin{equation}
\frac{1}{N^2 - 1}\, 
  \langle\langle
  {\rm Tr}\left[ W^{A\, \dagger}({\bf y})\, W^A({\bf x}) \right]
  \rangle\rangle_t 
  = \exp\left[-C_A\, v({\bf x}-{\bf y})\right]\, ,
  \label{3.9}
\end{equation}
or in the leading logarithmic approximation
\begin{equation}
\exp\left[-{({\bf x}-{\bf y})^2\over 8}{C_A\over C_F}Q^2_s\right]\, .
\end{equation}
With the explicit form of the Weizs\"acker-Williams field of the
quark projectile in configuration space,
\begin{equation}
        C_F\,   f({\bf x})\, f({\bf y}) = 
        \langle A_p(x)\, A_p(y) \rangle_{\rm proj} 
        = \frac{\alpha_s\, C_F}{2\pi} \, 
        \frac{{\bf x}\cdot {\bf y}}{{\bf x}^2\, {\bf y}^2}\, ,
       \label{3.10}
\end{equation}
the produced number spectrum (\ref{3.7}) for $q+A$ coincides with the 
expression given by Mueller and Kovchegov~\cite{al-yuri}.
Fourier transforming (\ref{3.7}), we see that it is 
nothing but the Gunion-Bertsch gluon radiation cross section~\cite{GB82} 
for a hard quark which receives a transverse momentum transfer
${\bf k}_x$ distributed with Gaussian probability distribution,
\begin{eqnarray}
  N_{\rm prod}({\bf k}) &=& \int d{\bf x}\, d{\bf y}\, e^{ik(x-y)}\,
        \frac{{\bf x}\cdot {\bf y}}{{\bf x}^2\, {\bf y}^2}
        \nonumber \\
        && \times
        \left( 1 + e^{-({\bf x}- {\bf y})^2\, {C_A\over 8C_F}Q_s^2}
        - e^{-{\bf x}^2\, {C_A\over 8C_F}Q_s^2} 
      - e^{- {\bf y}^2\,{C_A\over 8C_F} Q_s^2} \right)
    \nonumber \\
    &=& \frac{8\pi C_F}{C_A}\frac{1}{Q_s^2} \int d{\bf k}_x\, 
    e^{-{2C_F\over C_A Q_s^2}{\bf k}_x^2}\, 
    \frac{{\bf k}_x^2}{ {\bf k}^2\, 
    ({\bf k}_x - {\bf k})^2}\, .
    \label{3.11}
\end{eqnarray}

\subsection{Gluon production in $\pi(q\bar{q})$+$A$}
\label{sec3b}

As a second example, we consider for the initial state wave function
a quark-antiquark pair which may be viewed as a simplified representation
of a hadron. We take the quark to be at the transverse position ${\bf x}$ 
and the antiquark at ${\bf y}$. The state incoming into the interaction 
region is
\begin{eqnarray}
  U_-\, \vert (\alpha,{\bf x})(\bar\alpha,{\bf y})\rangle &\approx&      
  \vert (\alpha,{\bf x})(\bar\alpha,{\bf y})\rangle 
  + \int d{\bf z}\, \left [f({\bf z-x})\, 
  T^b_{\alpha\, \beta}\delta_{\bar\alpha \bar\beta} \right.
  \nonumber \\
  && \qquad \left. -
  f({\bf z-y})\delta_{\alpha\beta}\, T^{b}_{\bar\beta \bar\alpha}\right]\,
  \vert (\beta,{\bf x})\, (\bar\beta,{\bf y})\, (b,{\bf z})\rangle\, .
  \label{3.12}
\end{eqnarray}
Here, the colour indices of the quark and antiquark are independent, 
i.e., the incoming quark and antiquark is not taken to be in
a global colour singlet state.
The reason for this is the following. Any realistic hadron wave function 
apart from a quark and an antiquark has a large number of soft gluons.
Those are not the Weizs\"acker-Williams gluons present in eq.(\ref{3.12})
but rather genuine soft nonperturbative glue. They do not scatter eikonally
since their energy is small. One can reasonably expect that they are 
completely absorbed by the target. This absorption does not produce any 
gluons in the central rapidity region which we are attempting to
calculate here. However the presence of these soft gluons serves to randomize
the colour orientation of $q$ and $\bar q$.
We thus model the hadronic state as having all colour orientations of 
individual quark and antiquark as equally probable. The implication of this is
that the projection operator which we employ to project out the component
of the incoming hadron from the outgoing state, has independent summations
over the quark and antiquark colour indices.

The outgoing state is
\begin{eqnarray}
  \Psi_{out}&=&
  W^F_{\alpha\gamma}({\bf x})\, 
  W^{F\dagger}_{\bar\gamma\bar\alpha}({\bf y})\, \vert (\gamma,{\bf x})
  (\bar\gamma,{\bf y})\rangle 
  \nonumber \\
  && + \int d{\bf z}\, \left [f({\bf z}-{\bf x})\, 
  T^a_{\alpha\, \beta}W^F_{\beta\gamma}({\bf x})\,
  W^{F\dagger}_{\bar\gamma\bar\alpha}({\bf y})\, W^A_{ab}({\bf z})\, \right.
  \nonumber \\
  && \left. \quad -
  f({\bf z-y})T^{a}_{\bar\beta \bar\alpha}\, W^F_{\alpha\gamma}({\bf x})\,
  W^{F\dagger}_{\bar\gamma\bar\beta}({\bf y})\, W^A_{ab}({\bf z}) \right]\, 
  \vert (\gamma, {\bf x})\, (\bar\gamma,{\bf y})\, (b,{\bf z})\rangle\, ,
\label{3.13}
\end{eqnarray}
and after projecting out the initial state component

\begin{eqnarray}
  \delta\Psi_{\alpha\bar\alpha} &=& \int d{\bf z}\, 
  \Bigg [f({\bf z}-{\bf x})\, 
  T^a_{\alpha\, \beta}\, W^F_{\beta\gamma}({\bf x})\,
  W^{F\dagger}_{\bar\gamma\bar\alpha}({\bf y})\, W^A_{ab}({\bf z})\,
  \nonumber \\
  && \qquad\qquad -
  f({\bf z-y})T^{a}_{\bar\beta \bar\alpha}\, W^F_{\alpha\gamma}({\bf x})\,
  W^{F\dagger}_{\bar\gamma\bar\beta}({\bf y})\, W^A_{ab}({\bf z})
  \nonumber \\
  && \qquad\qquad - f({\bf z-x})\, 
  T^b_{\beta\gamma}\, W^F_{\alpha \beta}({\bf x})\,
  W^{F\dagger}_{\bar\gamma\bar\alpha}({\bf y})
  \nonumber \\
  && \qquad\qquad +
  f({\bf z-y})T^{b}_{\gamma\bar\beta}W^F_{\alpha\gamma}({\bf x})\,
  W^{F\dagger}_{\bar\beta\bar\alpha}({\bf y}) \Bigg ]\, 
  \vert (\gamma, x)\, (\bar\gamma,y)\, (b,{\bf z})\rangle\, .
\label{3.14}
\end{eqnarray}
Here again we have kept only terms of the lowest order in $f^2$.
The number of produced gluons is calculated in the state $\delta\Psi$ 
independently averaging over the initial state colour indices $\alpha$ and 
$\bar\alpha$. After some colour algebra we get
 \begin{eqnarray}
  &&N_{\rm prod}({\bf k}) = \frac{1}{N^2} \sum_{\alpha\bar\alpha}
  \langle \delta \Psi_{\alpha\bar\alpha}\vert a_d^\dagger({\bf k})\, 
  a_d({\bf k})\vert\, \delta\Psi_{\alpha\bar\alpha} \rangle
  \nonumber\\
  &&= C_F\, \int d{\bf z}\, d{\bf \bar z}\, e^{i{\bf k}\cdot({\bf z}-
  {\bf \bar z})}\, 
  \nonumber \\
  && \qquad \times \Bigg [\, f({\bf z-x})\, f({\bf \bar z-x})\, 
  \left( 1+ e^{-Nv({\bf z}-\bar{\bf z})}
                   - e^{-Nv({\bf z}-{\bf x})} 
                   - e^{-Nv(\bar{\bf z}-{\bf x})} \right) 
  \nonumber \\
  && \qquad   + f({\bf z-y})\, f({\bf \bar z-y})\, 
  \left( 1+ e^{-Nv({\bf z}-\bar{\bf z})}
         - e^{-Nv({\bf z}-{\bf y})}
           - e^{-Nv(\bar{\bf z}-{\bf y})}\right)\,
  \Bigg ]\, .
  \label{3.15}
\end{eqnarray} 
In comparison with the number spectrum (\ref{3.7}), 
(\ref{3.15}) is the independent sum of the radiation
spectra for two quark-nucleus collisions, with quarks at
transverse positions ${\bf x}$ and ${\bf y}$ respectively.
[This can be seen by using $v({\bf x}-{\bf y}) = 
\frac{({\bf x}-{\bf y})^2}{8\, C_F} Q_s^2$.]
In this sense, it appears sufficient to calculate the $q$-$A$
gluon number spectrum (\ref{3.7}) in order to determine
the $h$-$A$ number spectrum for ``hadrons'' $h$. However,
there are at least two obvious limitations to this simplified 
treatment: first, if colour correlations between the projectile 
partons become important, the gluon emission off different quarks 
is not independent any more. This will be seen in our discussion
of gluon production off a colour singlet $q\bar{q}$-Fock
state in section~\ref{sec4}. Second, if the colour charge density
in the projectile is large, then its WW field is not an
incoherent superposition of the WW fields of its constituents.
This in particular is important when both colliding objects
are nuclei, as in \cite{K00}.
We plan to address this question in a future publication \cite{inprep}.

\section{Small $x$ DIS in the target rest frame}
\label{sec4}

In section~\ref{sec3b}, we considered the propagation and
accompanying gluon radiation of a quark-antiquark pair in
a nuclear target without requiring that this colour dipole
is in a global colour singlet state. We argued that this
is a simple ansatz for a meson-nucleus collision where 
the hard quark and antiquark projectile components are not
required to be in a singlet state since soft non-perturbative 
gluon fields are present as well. The situation is different
for many problems in small $x$ DIS. In the nuclear target 
rest frame, the process of interest here is the eikonal
propagation of the hadronic Fock components of the virtual
photon through the nuclear target field. Due to formation
time arguments, these Fock states cannot develop a soft
colour-carrying field prior to the interaction. The
hadronic wavefunction of the virtual photon is hence perturbative,
and the lowest lying $q\bar{q}$ and $q\bar{q}g$ 
Fock states of the incoming virtual photon are in a colour
singlet. At not too low values of $x=x_0$ the photon wave function has only 
the $q\bar q$ component:
\begin{equation}
  \vert\gamma^*\rangle =\int d^2({\bf x-y})dz
\psi({\bf x- y},z) \frac{1}{\sqrt{N}} 
 \delta_{\alpha\, \bar{\alpha}}
                    |\alpha({\bf x})\, ,\bar{\alpha}({\bf y}),z\rangle\, .
\label{gammastar}
\end{equation}
Here as previously ${\bf x,y}$ are transverse coordinates of the quark and
antiquark respectively and $z$ is the fraction of the longitudinal momentum 
carried by the quark. The explicit form of the wave function $\psi$ 
for both transverse and longitudinal photon is given 
in many places (e.g. Ref.~\cite{yuri-larry}) and will not concern us here. 
We will also not indicate the $z$-dependence of the Fock states explicitly 
in the following, since the scattering amplitude of the $q\bar q$ pair in 
the eikonal approximation does not depend on $z$.

The DIS cross section is determined by the {\it total} cross section
of the $q\bar q$ pair on the target. The reason it is the {\it total} and 
not the {\it inelastic} probability that is important, is because even if 
only the phase of the $q\bar q$ state is changed in the scattering event, 
the coherence between the components of the $\gamma^*$ wave function is
disturbed and thus what emerges from the interaction region is some hadronic
state. The DIS cross section is thus given by
\begin{equation}
\sigma^{DIS}=\int d^2{\bf x}d^2{\bf y}dz
\psi({\bf x- y},z)\psi^*({\bf x- y},z)
P^{q\bar q}_{\rm tot}({\bf x},{\bf y})\, .
\label{sigmadis}
\end{equation}
At lower values of $x$ the virtual photon wave function develops a $q\bar q g$
Fock space component, due to the Weizs\"acker-Wiliams field of the 
quark and the antiquark. Thus the wave function of the $q\bar q$
pair in (\ref{gammastar}) is substituted by the normalized state
\begin{eqnarray}
&&\frac{1}{\sqrt{N}} 
 \delta_{\alpha\, \bar{\alpha}}
                    |\alpha({\bf x})\, ,\bar{\alpha}({\bf y})\rangle
 \rightarrow\Psi_{\rm in}=\frac{\cal N}{\sqrt{N}} 
\delta_{\alpha\, \bar{\alpha}}
                    |\alpha({\bf x})\, ,\bar{\alpha}({\bf y})\rangle
                    \nonumber \\
                  && \qquad \qquad \qquad \qquad \qquad \qquad 
                     + \frac{1}{\sqrt{N}} 
                      \int^{\ln{1/x}}_{\ln{1/x_0}}d\xi\int d^2{\bf z}
                      \left[ f({\bf z-x}) - f({\bf z-y}) \right]\, 
                      \nonumber \\
                  && \qquad \qquad \qquad \qquad \qquad \qquad \qquad 
                      \times T_{\alpha\, \beta}^a
                      |(\beta,{\bf x})\, (\alpha,{\bf y})\, 
                      (a,{\bf z},\xi)\rangle\, ,
  \label{4.1}
\end{eqnarray}
where in the leading logarithmic approximation we have
taken the gluon emission amplitude to be
independent of the gluon rapidity $\xi$.
This wavefunction is normalized to unity via the factor 
\begin{eqnarray}
  {\cal N} &=& \frac{1}{\sqrt{\left(1 + C_F\, \int d^2{\bf z}F^2({\bf x,y,z})
\ln(x_0/x)\right)}}\,\nonumber\\
           &\approx& 1 - \frac{C_F}{2}\, 
           \int d^2{\bf z}\, F^2({\bf x,y,z})\, \ln(x_0/x)\, .
\label{4.2}
\end{eqnarray}
Here and in what follows, we use the following notational shorthands 
for the Weizs\"acker-Williams fields,  
\begin{equation}
   F \equiv f({\bf z}-{\bf x}) - 
   f({\bf z}-{\bf y})\, .
  \label{4.3}
\end{equation}
The wave function (\ref{4.2}) is correct for Weizs\"acker-Williams 
photons with rapidity lower than that of both, the quark and the 
antiquark. The photons with intermediate rapidity can 
be emitted only by the fastest parton (be it $q$ {\it or} $\bar q$).
In the leading logarithmic approximation, however, it is the soft photons
that dominate the phase space and we only keep those in our analysis.

Even though only $P^{q\bar q}_{tot}$ is relevant for inclusive DIS 
cross section, in the rest of this section we will discuss also the 
inelastic and diffractive cross sections of the $q\bar q$ scattering 
on the nuclear target. Those have distinct physical meaning and
are useful for calculations of 
less inclusive quantities. Our starting point is 
the virtual photon component (\ref{4.1}).
The interaction with the target evolves  its wave function into
\begin{eqnarray}
 \Psi_{\rm out} &=& S\, \Psi_{\rm in}
                \nonumber \\
 &=& \frac{\cal N}{\sqrt{N}} 
     \left[ W^{F\, \dagger}({\bf y})\, W^F({\bf x})
          \right]_{\bar{\alpha}\, \alpha}
     |\alpha({\bf x})\, ,\bar{\alpha}({\bf y})\rangle
     \nonumber \\
 && + \frac{1}{\sqrt{N}}\, F\, 
     \left[ W^{F\, \dagger}({\bf y})\, T^a\, W^F({\bf x})
          \right]_{\bar{\alpha}\, \alpha}\, 
     W^A_{a\, b}({\bf z})\,  
     |\alpha({\bf x})\, ,\bar{\alpha}({\bf y})\, ,b({\bf z})\rangle \, ,
     \label{4.4}
\end{eqnarray}
where we have suppressed the integration over the gluon rapidity.

\subsection{$q\bar{q}$-contribution to photoabsorption cross sections}

We start with the discussion of the contributions from the lowest lying
$q\bar{q}$ Fock state. Details of the medium-average as well as
more differential cross sections are listed in the appendices.

The total scattering probability, which determines the inclusive DIS 
cross section according to (\ref{sigmadis}), reads
\begin{eqnarray}
  P_{\rm tot}^{q\bar{q}} &=& \,\langle\langle 2 -
  \frac{1}{N}
    {\rm Tr}\left[ W^F({\bf x})\, W^{F\dagger}({\bf y})\right] 
- \frac{1}{N}
    {\rm Tr}\left[ W^F({\bf y})\, W^{F\dagger}({\bf x})
\right] 
  \rangle\rangle 
                 \nonumber \\
   &=& 2\, \Bigg [\, 1 - e^{-C_F\, v({\bf x}-{\bf y})} \Bigg ]\, .
  \label{4.7}
\end{eqnarray}
For the inelastic photoabsorption cross section, we get
\begin{eqnarray}
  P^{q\bar q}_{\rm inel} &=& 1-\frac{1}{N^2}
   \langle\langle  {\rm Tr}\left[ W^{F\dagger}({\bf x})\, 
                   W^{F}({\bf y})\right]\rangle\rangle \,
   \langle\langle {\rm Tr}\left[ W^{F\dagger}({\bf y})\, 
                   W^{F}({\bf x})\right]\rangle\rangle
                 \nonumber \\
      &=& 1 - e^{-2C_F\, v({\bf x}-{\bf y})}\, ,
  \label{disinel}
\end{eqnarray}
and
\begin{eqnarray}
  P^{q\bar q\, {\rm proj}}_{\rm inel} &=& \langle\langle 1 - \frac{1}{N^2}
    {\rm Tr}\left[ W^{F\dagger}({\bf x})\, 
                   W^{F}({\bf y})\right] \,
    {\rm Tr}\left[ W^{F\dagger}({\bf y})\, 
                   W^{F}({\bf x})\right]\rangle\rangle
                 \nonumber \\
   &=& \langle\langle \frac{N^2-1}{N^2} - \frac{1}{N^2} W_{ab}^A({\bf x})
       W_{ab}^A({\bf y})  \rangle\rangle
   \nonumber \\ 
   &=& \frac{N^2-1}{N^2} \Bigg [\, 1 - e^{-N\, v({\bf x}-{\bf y})} \Bigg ]\, .
  \label{4.5}
\end{eqnarray}
As explained in section~\ref{sec2}, the difference between these two 
expressions is due to the target side diffractive processes which are not
included in (\ref{4.5}) but are accounted for in (\ref{disinel}).
Comparing (\ref{disinel}) and (\ref{4.5}) we see that
this difference is of the order $O(1/N^2)$.
The diffractive probability reads 
\begin{eqnarray}
  P_{\rm diff}^{q\bar{q}}
  &=& \langle\langle \left( \frac{1}{N}
    {\rm Tr}\left[ W^F({\bf x})\, W^{F\dagger}({\bf y})\right] 
    - 1\right)
    \left( \frac{1}{N}
    {\rm Tr}\left[ W^F({\bf y})\, W^{F\dagger}({\bf x})\right] 
    - 1\right) \rangle\rangle
  \nonumber \\
   &=& \langle\langle \frac{1}{N^2} \left(1 + W_{ab}^A({\bf x})
       W_{ab}^A({\bf y}) \right) 
    - \frac{1}{N} {\rm Tr}\left[ W^F({\bf x})\, W^{F\dagger}({\bf y})\right] 
    \nonumber \\
   && \quad 
    - \frac{1}{N} {\rm Tr}\left[ W^F({\bf y})\, W^{F\dagger}({\bf x})\right] 
    + 1 \rangle\rangle
   \nonumber \\ 
  &=& \left( \frac{N^2+1}{N^2} 
         - 2\, e^{-C_F\, v({\bf x}-{\bf y})} 
       +  \frac{N^2-1}{N^2} e^{-N\, 
          v({\bf x}-{\bf y})} \right)
   \nonumber\\
  &=& \left(1 - e^{-C_F\, v({\bf x}-{\bf y})} \right)^2
       + O\left(\frac{1}{N^2}\right)\, .
   \label{4.6}
\end{eqnarray}
The leading $O\left(\frac{1}{N}\right)$ contribution in
(\ref{4.6}) is the result given by Buchm\"uller et al. for the
diffractive photoabsorption probability \cite{buchmuller}. 
It is easy to see from eqs.(\ref{disinel},\ref{4.7},\ref{2.35}) 
that this contribution is precisely equal to 
the elastic probability $P_{el}$, since $P_{el}=P_{tot}-P_{inel}$.
This leading $O\left(\frac{1}{N}\right)$ result was later rederived 
in Ref.~\cite{yuri-larry} and it lead 
Kovchegov and McLerran to conjecture that the diffractive
cross section can be calculated by the procedure of averaging
over the target fields on the amplitude level. Physically,
however, as noted in Ref.~\cite{yuri-larry} and explained above,
such a procedure corresponds to (quasi)elastic 
scattering where each target colour source (nucleon) in the target nucleus
remains in a colour
singlet state. In contrast, the diffractive production requires
the outgoing projectile wavefunction to be in the colour singlet
state but it can leave some target sources in the colour octet and thus
the whole nucleus in an excited state (the
overall colour neutrality of the target is of course still preserved). 
Our derivation
makes it clear that for the diffractive cross section the averaging
must be done on the level of the cross section. 
We also see that the higher order in $1/N$ corrections 
have a clear physical meaning in this case - they distinguish
between the diffractive and the (quasi)elastic production.

As our previous discussion suggests the difference between the
diffractive and elastic cross sections for the scattering of
$q\bar q$ pair is due to precisely the same physical final states 
as the difference between $P_{\rm inel}$ and $P^{\rm proj}_{\rm inel}$.
Indeed subtracting from the total cross section (\ref{4.7})
the projectile biased inelastic one (\ref{4.5}), one obtains the diffractive 
cross section (\ref{4.6}).
For more complicated projectile
configurations this is not necessarily the case, since the global
color is not the only degree of freedom that characterizes the outgoing state.
This is also not true beyond the eikonal approximation where the transverse 
positions of the partons in the projectile may change during the interaction 
time \cite{Wdipole}.

\subsection{$q\bar{q}g$-Contributions to photoabsorption cross sections}
\label{sec4b}

The non-abelian Weizs\"acker-Williams field in the virtual photon
wavefunction (\ref{4.1}) is perturbatively small. In the calculations
of the last subsection, this was used to neglect all powers of $F$.
Now, we discuss the $O(F^2)$-corrections to these photoabsorption
cross sections. There are several motivations to do so. 
Though nominally suppressed by powers of the strong coupling
constant, such radiative corrections are known to give 
important contributions to the leading twist
photoabsorption cross sections in some regions
of phase space. Moreover, the gluon radiation off the $q\bar{q}$-Fock
state can be viewed as a leading contribution to the small $x$
evolution equations. This we shall discuss in section~\ref{sec5}. 
The $O(F^2)$-corrections arise from the $q\bar{q}g$-components
of the ingoing and outgoing $\gamma^*$ wavefunctions (\ref{4.1}),
(\ref{4.2}) and (\ref{4.4}). 

The total scattering probability reads
\begin{eqnarray}
&&P^{(q\bar q+q\bar q g)}_{\rm tot}=
\langle\langle \left(2 - \frac{1}{N}
    {\rm Tr}\left[ W^{F\dagger}({\bf x})\, 
                   W^{F}({\bf y})\right] 
  - \frac{1}{N}
    {\rm Tr}\left[ W^{F\dagger}({\bf y})\, 
                   W^{F}({\bf x})\right] 
 \right)\rangle\rangle
 \nonumber\\
    && \qquad + \ln({x_0/x})\,  F^2\,
\langle\langle \left\{\frac{1}{2}
    {\rm Tr}\left[ W^{F\dagger}({\bf x})\, 
                   W^{F}({\bf y})\right] 
  + \frac{1}{2}
    {\rm Tr}\left[ W^{F\dagger}({\bf y})\, 
                   W^{F}({\bf x})\right] \right.
  \nonumber\\
  &&\qquad\qquad\qquad\qquad\qquad 
  -{1\over 2N} {\rm Tr}\left[ W^{F\dagger}({\bf z})\, 
  W^{F}({\bf x})\right]  {\rm Tr}\left[ W^{F\dagger}({\bf y})\, 
                   W^{F}({\bf z})\right] 
  \nonumber\\
  &&\qquad\qquad\qquad\qquad\qquad 
  \left.-{1\over 2N} {\rm Tr}\left[ W^{F\dagger}({\bf x})\, 
  W^{F}({\bf z})\right]  {\rm Tr}\left[ W^{F\dagger}({\bf z})\, 
                   W^{F}({\bf y})\right] 
  \right\}\rangle\rangle\nonumber\\
&& \qquad = \left( 2 - 2e^{-C_Fv({\bf x}-{\bf y})}\right)
  \nonumber \\
&& \qquad + 2\, C_F\, \ln(x_0/x)
F^2\, \left( e^{-C_Fv({\bf x}-{\bf y})}
   - e^{-\frac{N}{2}[v({\bf z}-{\bf y})+v({\bf z}-{\bf x})]
        +\frac{1}{2N}v({\bf x}-{\bf y})}\right)\, .
\label{tot}
\end{eqnarray}
Here and in the following we use the shorthand
\begin{equation} 
F^2\equiv \int d^2{\bf z}F^2({\bf x,y,z})[...]\, ,
\end{equation} 
which stands for the operator acting by multiplication with the square 
of the WW field, and integration over its coordinate. Thus in this and the 
following expressions the coordinate of the gluon ${\bf z}$ is always 
integrated over.

The inelastic photoabsorption probability reads  
\begin{eqnarray}
  &&P^{(q\bar q+q\bar q g)}_{\rm inel}
  = 1- \frac{1}{N^2}
   \langle\langle{\rm Tr}\left[ W^{F\dagger}({\bf x})\, 
                   W^{F}({\bf y})\right] \rangle\rangle
  \langle\langle{\rm Tr}\left[ W^{F\dagger}({\bf y})\, 
                   W^{F}({\bf x})\right] \rangle\rangle
                 \nonumber\\
  &&\qquad +\ln(x_0/x) F^2 {1\over 2N^2}
  \langle\langle{\rm Tr}\left[ W^{F\dagger}({\bf y})\, 
                   W^{F}({\bf x})\right] \rangle\rangle
  \nonumber\\
  &&\times
 \langle\langle\Bigg\{N
{\rm Tr}\left[ W^{F\dagger}({\bf x}) W^{F}({\bf y})\right] 
 -{\rm Tr}\left[ W^{F\dagger}({\bf x}) W^{F}({\bf z})\right] 
{\rm Tr}\left[ W^{F\dagger}({\bf z}) W^{F}({\bf y})\right] 
\Bigg\}\rangle\rangle
\nonumber\\
&&\qquad  + \ln(x_0/x) F^2 {1\over2 N^2}
 \langle\langle{\rm Tr}\left[ W^{F\dagger}({\bf x})\, 
                   W^{F}({\bf y})\right] \rangle\rangle
\nonumber \\
&& \times \langle\langle\Bigg\{N
{\rm Tr}\left[ W^{F\dagger}({\bf y})\, 
                   W^{F}({\bf x})\right] 
 -{\rm Tr}\left[ W^{F\dagger}({\bf z}) W^{F}({\bf x})\right] 
{\rm Tr}\left[ W^{F\dagger}({\bf y}) W^{F}({\bf z})\right] \Bigg\}
\rangle\rangle
  \nonumber \\
  && = 1 - e^{-2C_Fv({\bf x}-{\bf y})}
     \nonumber \\
  && \quad + 2C_F  \ln(x_0/x)F^2 e^{-C_Fv({\bf x}-{\bf y})}
  \left(e^{-C_Fv({\bf x}-{\bf y})} -
        e^{-\frac{N}{2}[v({\bf z}-{\bf y}) + v({\bf z}-{\bf x})]
            + \frac{1}{2N} v({\bf x}-{\bf y})} \right)\, .
\label{inelqqg}
\end{eqnarray}
Finally, the diffractive probability reads
\begin{eqnarray}
   &&P_{\rm diff}^{(q\bar q+q\bar{q}g)} 
 = \langle\langle 
   \left( \frac{1}{N}
    {\rm Tr}\left[ W^F({\bf x})\, W^{F\dagger}({\bf y})\right] 
    - 1\right)
   \left( \frac{1}{N}
    {\rm Tr}\left[ W^F({\bf y})\, W^{F\dagger}({\bf x})\right] 
    - 1\right)
    \nonumber \\
   && \qquad +  C_F\,\ln(x_0/x) F^2\, \left\{
    -\left( \frac{1}{N}
    {\rm Tr}\left[ W^F({\bf x})\, W^{F\dagger}({\bf y})\right] 
    - 1\right) \right.
   \nonumber \\
    && \qquad \qquad \qquad \qquad \qquad  \times \left( \frac{1}{N}
    {\rm Tr}\left[ W^F({\bf y})\, W^{F\dagger}({\bf x})\right] 
    - 1\right) \nonumber\\
&&  \qquad \qquad \qquad \qquad +\left( \frac{1}{N^2-1}
     {\rm Tr}\left[ W^{F\dagger}({\bf y}) W^F({\bf z})\right]\, 
     {\rm Tr}\left[ W^{F\dagger}({\bf z}) W^F({\bf x})\right]\,
      \right.
     \nonumber \\
&&\left.  \qquad \qquad \qquad \qquad \qquad 
      -\frac{1}{N(N^2-1)}
    {\rm Tr}\left[ W^F({\bf x})\, W^{F\dagger}({\bf y})\right] 
    - 1\right)
   \nonumber \\
&& \quad \qquad \qquad  \qquad \qquad \times \left( \frac{1}{N^2-1}
     {\rm Tr}\left[ W^{F\dagger}({\bf x}) W^F({\bf z})\right]\, 
     {\rm Tr}\left[ W^{F\dagger}({\bf z}) W^F({\bf y})\right]\,
      \right.
     \nonumber \\
&&\left. \left.  \qquad \qquad \qquad \qquad \qquad 
      -\frac{1}{N(N^2-1)}
    {\rm Tr}\left[ W^F({\bf y})\, W^{F\dagger}({\bf x})\right] 
    - 1\right)\right\}\rangle\rangle\, .
\label{4.14}
\end{eqnarray}
The explicit averaging of this expression over the target field 
is calculated in Appendix~\ref{appc}. The final result is
\begin{eqnarray}
   &&P_{\rm diff}^{(q\bar q+q\bar{q}g)}  = \left( \frac{N^2+1}{N^2} 
         - 2\, e^{-C_F\, v({\bf x}-{\bf y})} 
       +  \frac{N^2-1}{N^2} e^{-N\, 
          v({\bf x}-{\bf y})} \right)
      \nonumber \\
     &&+ C_F\, \ln(x_0/x)F^2\, \Bigg [\, -  \frac{N^2-2}{N^2(N^2-1)} 
         +2\, e^{-C_F\, v({\bf x}-{\bf y})} 
       -  \frac{(N^2-1)^3-1}{N^2(N^2-1)^2} e^{-N\, 
          v({\bf x}-{\bf y})}\nonumber\\
&&+\frac{N^2-2}{N^2(N^2-1)}
 \left(e^{-Nv({\bf z}-{\bf x})} + e^{-Nv({\bf z}-{\bf y})}\right)
+\frac{N^2-4}{2(N^2-1)}e^{-N[v({\bf z}-{\bf x})
         + v({\bf z}-{\bf y})]} 
\nonumber \\
      &&- 2e^{-\frac{N}{2}[v({\bf z}-{\bf x})
         + v({\bf z}-{\bf y})]}\left(e^{ \frac{1}{2N}v({\bf x}-{\bf y})}
-\frac{2}{N^2(N^2-1)}e^{ -\frac{N}{2}v({\bf x}-{\bf y})}\right)
       \nonumber \\
&&+\frac{N^2}{4(N^2-1)}\left(\frac{N+3}{N+1}e^{-(N+1)[v({\bf z}-{\bf x})
         + v({\bf  z}-{\bf y})] + v({\bf x}-{\bf y})} \right.
\nonumber \\
&& \qquad \qquad \qquad \qquad 
   \left. +\frac{N-3}{N-1}e^{-(N-1)[v({\bf z}-{\bf x})
         + v({\bf z}-{\bf y})] - v({\bf x}-{\bf y})}\right)\Bigg ]\, .
\label{4.14a}
\end{eqnarray}
To leading order in $1/N$ this simplifies to
\begin{eqnarray}
   &&P_{\rm diff}^{(q\bar q+q\bar{q}g)} = 
   \left(1 - e^{-\frac{N}{2}v({\bf x}-{\bf y})}\right)^2\,
      \nonumber \\
      && \qquad + C_F\, F^2\, \ln(x_0/x)\,  \left\{
     -  \left(1 - e^{-\frac{N}{2}v({\bf x}-{\bf y})}\right)^2\,
     + \left( 1 - e^{-\frac{N}{2}[v({\bf z}-{\bf x}) + v({\bf z}-{\bf y})]}
          \right)^2\right\}\nonumber\\
&& \qquad + O\left(\frac{1}{N^2}\right)\, .
\label{4.14b}
\end{eqnarray}
For completeness, we also give the projectile biased inelastic
probability
\begin{eqnarray}
  &&P_{\rm inel}^{(q\bar q+q\bar{q}g)\, {\rm proj}} 
  = \langle\langle
  1 - {\cal N}^4\left( \frac{1}{N^2} + \frac{1}{N^2} W_{ab}^A({\bf x})
       W_{ab}^A({\bf y})\right)
     \nonumber \\
   && \quad \qquad - \frac{2}{N^2}  \ln(x_0/x)F^2  \Bigg [\,  
       \frac{1}{2N} W_{ab}^A({\bf z}) W_{ab}^A({\bf x})
      +  \frac{1}{2N} W_{ab}^A({\bf z}) W_{ab}^A({\bf y})
    \nonumber \\
   && \quad \qquad \quad 
      + \frac{1}{4} W_{ab}^A({\bf z})\, W_{cd}^A({\bf y})\, 
    W_{ef}^A({\bf x})\, 
    \left( d_{ace}d_{bdf} + f_{ace}f_{bdf}\right)  \Bigg ]\, \rangle\rangle
  \nonumber \\
  && \qquad
   = \frac{N^2-1}{N^2} \left(1-e^{-Nv({\bf x}-{\bf y})}\right)
  \nonumber \\
  && \quad \qquad + 2C_F  \ln(x_0/x)F^2 \Bigg \lbrace \frac{1}{N^2} 
     + \frac{N^2-1}{N^2}e^{-Nv({\bf x}-{\bf y})}
     - \frac{1}{N^2}e^{-Nv({\bf z}-{\bf x})}
   \nonumber \\
  && \qquad \qquad 
     - \frac{1}{N^2}e^{-Nv({\bf z}-{\bf y})} 
     - \left(1-\frac{2}{N^2}\right)\, 
     e^{-\frac{N}{2}\, [v({\bf z}-{\bf x}) + v({\bf z}-{\bf y})
                      + v({\bf x}-{\bf y})]} \Bigg \rbrace\, .
   \label{4.8}
\end{eqnarray}
To leading order $O\left(\frac{1}{N}\right)$, the $q\bar{q}$
contribution to the projectile biased
and true photoabsorption cross sections coincide. However, for
the $q\bar q g$ configuration, the elastic and the 
diffractive  cross sections do not agree
even at the leading order in $1/N$. The elastic cross section can be 
immediately inferred from (\ref{tot}). To leading order in $1/N$
\begin{eqnarray}
  &&P_{\rm el}^{(q\bar q+q\bar qg)} = \left( 
  1 - e^{-{N\over 2}v({\bf x}-{\bf y})}\right)^2
  \nonumber\\
  && \qquad + 2C_F\, \ln(x_0/x)F^2\,\left( 
  1 - e^{-{N\over 2}v({\bf x}-{\bf y})}\right) 
  \left( e^{-{N\over 2}v({\bf x}-{\bf y})}
   - e^{-\frac{N}{2}[v({\bf z}-{\bf y})+v({\bf z}-{\bf x})]}
       \right)\nonumber\\
 && \qquad +O(1/N^2)\, .
\end{eqnarray}
This has to be compared with (\ref{4.14b}). The difference in the $O(F^2)$ 
terms is due to the fact that as opposed to the $q\bar q$ case, the 
quantum state now is characterized by a new degree of freedom. This degree
of freedom is the 
relative phase between the $q\bar q$ and the $q\bar qg$ components of the
wave function.
Thus the outgoing state can be a colour singlet which is not identical to the 
incoming one.

\section{Eikonal low x evolution}
\label{sec5}

The results of the previous section allow us to derive evolution 
equations for inelastic and diffractive DIS cross sections in the 
eikonal approximation. In the operator form, the equation for the 
evolution of the inelastic cross section is the same as that of 
Balitsky~\cite{bk}. Our derivation is very simple and allows us 
to trace explicitly some of the effects that limit the range of 
validity of this equation. We also present  
the all order $1/N$-corrections to
Kovchegov's version of this evolution equation. Those are obtained 
with the use of Fierz identities as described in the appendices.
The equation for the diffractive 
cross section as given here in the operator form is new.
In the large N limit it reduces to the equation discussed in 
Ref.~\cite{yuri-genya} and we present $1/N$ corrections to this form.

\subsection{The inelastic cross section}
\label{sec5a}

Consider DIS at some low value of $x_0$. We work in the frame where 
the photon wave function
at $x_0$ has only a $\bar q q$ component, but the target ensemble is
characterised by large values of the field. There are two distinct physical
situations that correspond to this. One is the scattering 
on a nuclear target at not very high energy. In this case the large target
fields are due to large number of nucleons in the target rest frame, and
no significant $x$ evolution has taken place ($x_0$ is not very low).
Another situation is that the scattering is on a single hadron and $x_0$ is 
very small. In this case the large target fields have been generated due to
the low $x$ evolution which in our frame has been put into the target.
There are some clear physical differences between the two situations, 
which we will discuss later, but for now we treat them both in the same
framework.

Thus at $x_0$ the DIS cross section is given by (\ref{sigmadis}).
The first line of (\ref{4.7}) then expresses it in terms of the
glue fields of the target, while the second line gives its explicit 
expression in terms of our ansatz for the target field averages.
Now consider a lower value of $x$. It is convenient 
to use the frame where all the evolution between $x_0$ and $x$
is occuring in the photon wave function. 
This means that to achieve higher interaction energy the photon is boosted,
while the target remains the same.
Thus the target field ensemble is
unchanged, but the photon wave function at $x$ has an extra $\bar q q g$ 
component according to (\ref{4.1}) and (\ref{4.2}).
The inelastic scattering probability is now given according to 
(\ref{sigmadis}) but with $P^{(q\bar q+q\bar qg)}$ 
of (\ref{tot}) substituted for $P^{q\bar q}$.
If we were to put the 
evolution in the target instead, the inelastic cross section would still be 
given by (\ref{sigmadis}) but with the average performed over the new 
target field ensemble. We can thus combine (\ref{4.7}) and (\ref{tot}) into 
a differential equation 
governing the change of the target field averages with $x$ 
(we define $\xi\equiv\ln(x_0/x)$)
\begin{eqnarray}
&&-{d\over d \xi}
   \langle\langle {\rm Tr}\left[ W^{F\dagger}({\bf x})\, 
     W^{F}({\bf y})\right] \rangle\rangle
   =F^2\langle\langle  {N\over 2}
   {\rm Tr}\left[ W^{F\dagger}({\bf x}) W^{F}({\bf y})\right] 
\nonumber \\
&& \qquad \qquad \qquad \qquad\qquad
   - {1\over 2}
   {\rm Tr}\left[ W^{F\dagger}({\bf x}) W^{F}({\bf z})\right] 
   {\rm Tr}\left[ W^{F\dagger}({\bf z}) W^{F}({\bf y})\right] 
\rangle\rangle\, .
\label{bk}
\end{eqnarray}
This equation is identical to that derived by Balitsky~\cite{bk}.
Taking now the averages over the target fields and 
denoting the ``dipole scattering probability'' by
\begin{equation}
  D({\bf x},{\bf y})
  = 1 - e^{-C_F\, v({\bf x},{\bf y})}\, , 
  \label{5.2}
\end{equation}
we have
\begin{eqnarray}
 &&{d\over d\xi}D({\bf x},{\bf y})=C_FF^2 \Bigg\lbrace
   1-D({\bf x},{\bf y})
   \nonumber \\
   && \qquad \qquad - [1-D({\bf z},{\bf y})]^{N^2\over N^2-1}
             [1-D({\bf x},{\bf z})]^{N^2\over N^2-1}
             [1-D({\bf x},{\bf y})]^{1\over N^2-1}\Bigg\rbrace\, .
\label{D}
\end{eqnarray}
To leading order in $1/N$ and assuming that
$D$ does not depend on the impact parameter ($b=(x+y)/2$) so long as
$b$ is smaller than the target radius $R$,
this reduces to the equation derived by Kovchegov~\cite{bk}. 

The extra terms in (\ref{D}) furnish  
$1/N$ corrections to Kovchegov's form of the evolution equation.
Interestingly enough the corrections arise as noninteger powers of $[1-D]$, 
and thus at small $D$ correspond to infinite series. In the saturation 
regime, where $D$ is close to unity the effect of these corrections may 
very well be large, and it would be quite interesting to explore these 
effects numerically, similarly to Ref.~\cite{motyka}.

These corrections have a very simple meaning. As discussed by 
Kovchegov~\cite{bk}, the leading $1/N$ approximation in the present 
context is equivalent to assuming that the dipoles interact with the 
target independently. From our derivation of the target averages in 
Appendix~\ref{appc}, it is obvious that the corrections in (\ref{D}) 
arise precisely due to the breakdown of the independent propagation
beyond the leading order in $1/N$.

What is meant by ``independent propagation''? Physically, as discussed 
in section~\ref{sec2}, in the eikonal approximation the partons of the 
projectile propagate through the target field completely independently 
of each other. This statement follows from the eikonal kinematics and
does not rely on the $1/N$ expansion. However the fields in the target 
are correlated with each other at different transverse coordinates. As 
a result the eikonal phases picked up by the partons are not independent 
when averaged over the target field ensemble. If the target fields are 
small locally at every longitudinal coordinate, so that our model for 
averaging is applicable, then in the leading order in $1/N$ all eikonal 
phase averages factorize into amplitudes corresponding to fundamental 
dipoles. In this sense, the dipoles ``propagate independently'' only
to leading order in $1/N$. Moreover, it is crucial for this
factorization that any given source in the target can exchange at most
two gluons with the projectile. If the fields are locally not small 
it is possible that the propagation is still eikonal, but the 
$1/N$ factorization breaks down. 

We finally discuss the limitations of this approach.
There are at least two reasons to expect that the evolution
equation derived here breaks down at very low values of $x$:

Firstly, the derivation assumes that all the evolution up until $x_0$ 
occured in the target wave function. Further evolution is nothing but 
boosting of the target. In the boosted frame
the gluon fields contract in the longitudinal direction and their amplitude
grows, as discussed in Refs.~\cite{mv,jklw}. 
This does not mean that the effective width
of the target shrinks, since the low momentum components of the glue
field also become important. The effective width of the target then
can remain approximately constant, but the fields inside it grow.
Therefore if the target wave function has undergone a lengthy low $x$ 
evolution, the target fields have significantly increased locally and
the averaging procedure employed here and in Ref.~\cite{bk} breaks down.
Thus we expect (\ref{D}) to be a sensible description for targets which
are large but ``soft'', like a large nucleus in its rest frame, but not for
targets which are ``hard'', like a very energetic proton.

Secondly, if the evolution is followed to very small values of $x$,
the projectile wave function itself becomes dense. Then any further 
evolution has to take into account that the Weizs\"acker-Williams field 
of the {\it projectile} becomes large. Further evolution then becomes 
nonlinear already on the projectile side, since we would not be able 
to neglect higher orders of the field $F$ in all our calculations. 
The onset of this effect may be expected to occur at the scale
\begin{equation}
  \ln (x_0/x) \simeq \frac{1}{F^2}\, ,
\end{equation}
when the $q\bar{q}$ and $q\bar{q}g$ contributions to the photoabsorption
cross section become comparable. In the language of the 
second paper in Ref.~\cite{bk}, such higher $O(F^2)$-corrections
are due to ``Pomeron loops''. The evolution equation then breaks 
down already in the operatorial form. 
Another way of saying it, is that the averages of the products of large
number of Wilson lines is not described by a straightforward generalization
of (\ref{bk}) given in Ref.~\cite{bk}. 
It is an interesting question whether the evolution equation of \cite{jklw}
accounts better for this physics. As discussed in \cite{guilherme}, 
the JKLW evolution is different form the BK evolution, when the 
fluctuation fields are large. This is precisely the case of a large projectile.
However whether this difference is due to different physics content or to
technical difficulties with JKLW equation
is not clear at the moment\footnote{
It has been recently suggested\cite{larry} that the apparent differences 
between the BK and JKLW equations are due to the incorrect derivation of 
the latter. However, since the discussion of \cite{larry} does not 
point to a mistake in the calculations of \cite{jklw,guilherme}, 
we consider the question open as 
discussed in \cite{guilherme}.}, as discussed in \cite{guilherme}.

\subsection{Diffractive cross section}
\label{sec5b}
We can repeat the same excercise for the diffractive cross section.
Using eqs.(\ref{4.6}) and (\ref{4.14}) we get
\begin{eqnarray}
\label{opdif}
&&{d\over d\xi}P_{\rm diff}(x,y)= -C_FF^2\,
    \langle\langle
    \left( \frac{1}{N}
    {\rm Tr}\left[ W^F({\bf x})\, W^{F\dagger}({\bf y})\right] 
    - 1\right)
  \nonumber \\
   && \qquad\qquad\qquad\qquad\qquad\qquad\times \left( \frac{1}{N}
    {\rm Tr}\left[ W^F({\bf y})\, W^{F\dagger}({\bf x})\right] 
    - 1\right)
    \nonumber \\
&& \qquad \qquad \qquad\qquad - \left( \frac{1}{N^2-1}
     {\rm Tr}\left[ W^{F\dagger}({\bf y}) W^F({\bf z})\right]\, 
     {\rm Tr}\left[ W^{F\dagger}({\bf z}) W^F({\bf x})\right]
     \right. \nonumber \\
&& \qquad \qquad\qquad\qquad \qquad\left. -\frac{1}{N(N^2-1)}
    {\rm Tr}\left[ W^F({\bf x})\, W^{F\dagger}({\bf y})\right] 
    - 1\right) \nonumber\\
&& \qquad \qquad \qquad\qquad\quad \times \left( \frac{1}{N^2-1}
     {\rm Tr}\left[ W^{F\dagger}({\bf x}) W^F({\bf z})\right]\, 
     {\rm Tr}\left[ W^{F\dagger}({\bf z}) W^F({\bf y})\right]
     \right. \nonumber \\
&& \qquad \qquad \qquad \qquad \qquad \quad\left. -\frac{1}{N(N^2-1)}
    {\rm Tr}\left[ W^F({\bf y})\, W^{F\dagger}({\bf x})\right] 
    - 1\right) \rangle\rangle\, .
    \label{5.5}
\end{eqnarray}
This is the operatorial form of the low $x$ evolution equation for
the diffractive photoabsorption cross section. 

A comment is in order about the precise meaning of $P_{\rm diff}$.
We have defined $P_{\rm diff}$ at $x_0$ as the probability that the $q \bar q$ 
pair in the photon wave function remains in a colour singlet state after
the interaction with the target. At lower value of $x$ the wave function
contains a quark-antiquark-gluon component. 
The rapidity of this Weizs\"aker-Williams
gluon in the leading logarithmic approximation is always 
smaller than the rapidity of the quark or antiquark 
but greater than the initial rapidity $\ln (1/x_0)$. Again we require
that the outgoing state is in the colour singlet. This requirement
does not constrain the emission into the
rapidity interval between the gluon and the quark-antiquark pair in the final
state, but it does not allow emission of particles with rapidity smaller
than that of the softest gluon, that is $\ln (1/x_0)$. Thus the final states
allowed by our definition have rapidity gap on the target side of at least
$\ln(1/x_0)$.

Our definition of $P_{\rm diff}$ coincides therefore with that 
of Ref.~\cite{yuri-genya} and (\ref{opdif}) is the operator form of the
evolution equation for this quantity. It is also easy to see that in the
leading order in $1/N$ this equation indeed coincides with the equation
derived in Ref.~\cite{yuri-genya}. The straightforward way is again to realize
that in the leading order in $1/N$ the target field averages factorize. Thus
the right hand side of (\ref{opdif}) can be factorized into products of 
$P_{\rm diff}=\langle\langle\left( \frac{1}{N}
    {\rm Tr}\left[ W^F({\bf x})\, W^{F\dagger}({\bf y})\right] 
    - 1\right)\left( {\rm h.c.} \right)
\rangle\rangle$ (denoted  in \cite{yuri-genya} by $N_D$) and
$P_{\rm tot}=\langle\langle\left( 1-\frac{1}{N}
    {\rm Tr}\left[ W^F({\bf x})\, W^{F\dagger}({\bf y})\right]
\right)\rangle\rangle$ (denoted  in \cite{yuri-genya} by $N_0$).
Neglecting the explicit $1/N^2$ corrections in the coefficients of 
(\ref{opdif}) the result is
identical to the differential form of eq.(11) of \cite{yuri-genya}.
\begin{eqnarray}
\label{opdifN}
&&{d\over d\xi}P_{\rm diff}(x,y)= C_FF^2\,
\{-P_{\rm diff}(x,y) -P_{\rm diff}(x,z) -P_{\rm diff}(y,z)
\nonumber \\
&&\qquad \qquad \qquad +P_{\rm diff}(x,z)P_{\rm diff}(z,y)
-2P_{\rm diff}(x,z)P_{\rm tot}(z,y)\nonumber\\
&&\qquad \qquad \qquad-2P_{\rm tot}(x,z)P_{\rm diff}(z,y)
+2P_{\rm tot}(x,z)P_{\rm tot}(z,y)\}\, .
\end{eqnarray}

We note however that within the region of validity of the present 
approximation,
(\ref{opdifN}) is not really a differential equation, but it rather
determines $P_{\rm diff}$ directly in terms of the dipole cross 
scattering probability  $D$ of (\ref{5.2}). The point is that the 
right hand side of (\ref{opdif}) in the leading order in $1/N$
can be written entirely in terms of $D$:
\begin{equation}
\label{opdifN1}
{d\over d\xi}P_{\rm diff}(x,y)=  C_FF^2\,
\left\{-D^2(x,y)+\left[D(x,z)D(z,y)-D(x,z)-D(z,y)\right]^2\right\}\, .
\end{equation}
Thus once $D(\xi)$ 
is found by solving (\ref{D}), the diffractive probability
can be found by directly integrating the right hand side of (\ref{opdifN1}).
The reason (\ref{opdifN1}) is consistent with (\ref{opdifN}) is the 
following. The initial condition satisfied by $P_{\rm diff}$ as discussed in
\cite{yuri-genya} is 
\begin{equation}
P_{\rm diff}(x=x_0)=D(x_0)^2\, .
\end{equation}
At lower values of $x$ the solution of (\ref{opdifN1}) of course will 
not satisfy this relation any longer. However 
the difference between $P_{\rm diff}$ and $D^2$ at any value of $x$ is small,
of order $\ln (x_0/x)F^2$. Thus in the right hand side of the differential 
equation (\ref{opdifN}) the two can be interchanged as long as
$\ln(x_0/x)\ll 1/F^2$. As discussed in the previous subsection,
this condition has to be satisfied anyway for the whole scheme to be 
consistent. For even lower values of $x$ the 
corrections nonlinear in $F^2$ (or
``Pomeron loop'' diagrams) become
important, and those are not accounted for in the present approximation 
(as well as in the derivations of Refs.~\cite{bk,yuri-genya}).

Thus we stress again, that in order to find the diffractive probability 
one does
not have to solve an independent evolution equation, but rather integrate
(\ref{opdifN1}) with the known right hand side. This statement remains true
beyond the leading order in $1/N$. The complete set of $1/N$ corrections
due to correlated dipole propagation is read off eq.(\ref{4.14a}),
\begin{eqnarray}
&&{d\over d\xi}P_{\rm diff}(x,y)= C_FF^2\, 
    \Bigg [\, -  \frac{N^2-2}{N^2(N^2-1)} 
         +2\, e^{-C_F\, v({\bf x}-{\bf y})} 
       -  \frac{(N^2-1)^3-1}{N^2(N^2-1)^2} e^{-N\, 
          v({\bf x}-{\bf y})}\nonumber\\
&&+\frac{N^2-2}{N^2(N^2-1)}
 \left(e^{-Nv({\bf z}-{\bf x})} + e^{-Nv({\bf z}-{\bf y})}\right)
+\frac{N^2-4}{2(N^2-1)}e^{-N[v({\bf z}-{\bf x})
         + v({\bf z}-{\bf y})]} 
\nonumber \\
      &&- 2e^{-\frac{N}{2}[v({\bf z}-{\bf x})
         + v({\bf z}-{\bf y})]}\left(e^{ \frac{1}{2N}v({\bf x}-{\bf y})}
-\frac{2}{N^2(N^2-1)}e^{ -\frac{N}{2}v({\bf x}-{\bf y})}\right)
       \nonumber \\
&&+\frac{N^2}{4(N^2-1)}\left(\frac{N+3}{N+1}e^{-(N+1)[v({\bf z}-{\bf x})
         + v({\bf z}-{\bf y})] + v({\bf x}-{\bf y})}
    \right. \nonumber \\
&& \qquad \qquad \left. +
  \frac{N-3}{N-1}e^{-(N-1)[v({\bf z}-{\bf x})
         + v({\bf z}-{\bf y})] - v({\bf x}-{\bf y})}\right)\Bigg ]\, .
 \label{5.8}
\end{eqnarray}
Here, $v({\bf x-y})$ is related to the total dipole cross section 
$D({\bf x-y})$ via (\ref{5.2}).

\section{Summary}

In this paper we have discussed in a very explicit and simple way 
the physics of high energy processes in the eikonal approximation.
This provides a simple unifying framework for a variety of calculations
existing in the literature. In particular we 
considered processes where a small projectile (a hadron or a virtual
photon) scatters on a large target (a nucleus).
Our approach then allowed us to derive the inclusive one particle spectrum
as well as total, inelastic and diffractive cross sections. In particular
the low x evolution equation for the DIS cross section~\cite{bk}
follows immediately from this calculation. 
The various sources of the breakdown of this evolution equation at very
low $x$ are also seen clearly in our derivation.
We have also derived
the operatorial form for the evolution of the diffractive cross section
and have shown that in the large $1/N$ limit it reduces to the equation
of Ref.~\cite{yuri-genya}.
We have clarified that this evolution is not independent 
of the total photoabsorption cross section, and indeed
the diffractive cross section is determined by a direct integration once the
total dipole scattering probability  (at fixed impact parameter) is known.

The averaging over the target nucleus wave function has been performed
in the approximation where the target is considered to be a dilute system of
weak sources of the gluon field. We did however calculate
explicitly all-order $1/N$ corrections in this model.
This allowed us to provide all-order corrections
to the various cross sections as well as to 
both evolution equations. These corrections stem from the fact that 
fundamental dipoles
do not propagate through the target independently from each other.
These corrections have not appeared in the literature so far.

We hope that this work is useful in providing a simple explicit 
unifying framework for the eikonal physics. The method discussed 
here allows simple calcutions of many other observables at high energy. 
What we consider to be most interesting are less inclusive observables
giving detailed information about the structure of the final states.
It is for example straightforward
to determine evolution of such quantities as the inclusive one particle
distributions and cross sections for diffractive vector meson production
\cite{inprep}.

{\bf Acknowledgement:} We thank Yuri Kovchegov and Sangyong Jeon for
interesting discussions and for pointing out typos in an earlier
version of this manuscript.

\appendix
\section{Photoabsorption cross sections}
\label{appa}

In this appendix, we give full expressions for the differential
inelastic ($c = {\rm inel}$), diffractive ($c = {\rm diff}$) and 
total ($c = {\rm tot}$) photoabsorption cross sections. We consider
contributions from the $q\bar{q}$- and $q\bar{q}g$- Fock 
components of the virtual photon. We also give expressions for 
the differential cross sections with respect to the
transverse momenta ${\bf p}_1$, ${\bf p}_2$ and ${\bf k}$ of the
quark, antiquark and gluon respectively. The calculation
of the medium average $\langle\langle \dots \rangle\rangle$ will be
discussed in Appendix~\ref{appc}. 

The wavefunctions $\delta \Psi_c$
are constructed as discussed in section~\ref{sec2}. 

The $S$-matrix element, or overlap (\ref{2.6})
at fixed transverse coordinates reads
\begin{eqnarray}
  &&\langle \Psi^{(q\bar q+q\bar qg)}_{\rm in}({\bf x,y})| 
  \Psi^{(q\bar q+q\bar qg)}_{\rm out}({\bf x,y})\rangle 
  =\langle\langle\frac{{\cal N}^2}{N}\, {\rm Tr}\left[ W^{F\dagger}({\bf y})\, 
    W^F({\bf x})\right] \nonumber \\
&&\qquad \qquad + \frac{1}{N}
 \ln(x_0/x)\int d^2{\bf z} \vert f({\bf z}-{\bf x}) 
 - f({\bf z}-{\bf y}) \vert^2
 \nonumber \\
&& \qquad \qquad \qquad \qquad\qquad \qquad \times
 {\rm Tr}\left[ T^a\, W^F({\bf x})\, T^b\, W^{F\dagger}({\bf y})\right]\,
  W_{ab}^A({\bf z})\rangle\rangle\, .
  \label{a.3} 
\end{eqnarray}
Treating higher powers of the WW fields $f$ in the normalization factor 
${\cal N}$
as perturbatively small contributions which can be neglected, and using 
(\ref{b.5}), one finds
\begin{eqnarray}
  &&\langle \Psi^{(q\bar q+q\bar qg)}_{\rm in}({\bf x,y})| 
  \Psi^{(q\bar q+q\bar qg)}_{\rm out}({\bf x,y})\rangle 
  =\langle\langle\frac{1}{N}\, {\rm Tr}\left[ W^{F\dagger}({\bf y})\, 
    W^F({\bf x})\right] \\
&&+ \frac{1}{2N}
  \ln(x_0/x)\int d^2{\bf z} \vert f({\bf z}-{\bf x}) 
                    - f({\bf z}-{\bf y}) \vert^2\nonumber\\
&&\times\Bigg \lbrace
 {\rm Tr}\left[ W^{F\dagger}({\bf z})\, 
    W^F({\bf x})\right]  {\rm Tr}\left[ W^{F\dagger}({\bf y})\, 
    W^F({\bf z})\right] - N{\rm Tr}\left[ W^{F\dagger}({\bf y})\, 
    W^F({\bf x})\right] \Bigg \rbrace \rangle\rangle\, .\nonumber
  \label{a.3a} 
\end{eqnarray}
For the diffractive cross section (\ref{2.17}), 
we use the projection operator on the colour singlet states 
\begin{eqnarray}
  {\cal P}_{\rm singlet}^{q\bar{q}} &=& \frac{1}{N}
           \delta_{\alpha\, \bar{\alpha}} 
                  |\alpha\, ,\bar{\alpha}\rangle
           \langle \beta\, ,\bar{\beta} | \delta_{\beta\, \bar{\beta}}\, ,
           \label{a.5}\\
  {\cal P}_{\rm singlet}^{q\bar{q}g} &=&
           \frac{2}{N^2 - 1} 
           T_{\alpha\, \beta}^a |\alpha\, \beta\, a\rangle
           \langle \bar{\alpha}\, \bar{\beta}\, \bar{a}|  
           T_{\bar{\alpha}\, \bar{\beta}}^{\bar{a}}\, .
           \label{a.6}
\end{eqnarray}
We start with the contribution from the $q\bar{q}$ component.
The contribution to the total cross section reads
\begin{eqnarray}
  &&\langle\delta\Psi^{q\bar{q}}_{\rm tot}({\bf x},{\bf y})\,\vert 
        \delta\Psi^{q\bar{q}}_{\rm tot}({\bf x},{\bf y})\rangle
        = \nonumber\\
  && \qquad \langle\langle2-  \frac{1}{N}
    {\rm Tr}\left[ W^{F\dagger}({\bf x})\, 
                   W^{F}({\bf y})\right] 
    - \frac{1}{N} {\rm Tr}\left[ W^{F\dagger}({\bf y})\, 
                   W^{F}({\bf x})\right]\rangle\rangle  \, .
   \label{a.8}
\end{eqnarray}
For the inelastic cross section we have
\begin{eqnarray}
  &&\langle\delta\Psi^{q\bar{q}}_{\rm inel}({\bf x},{\bf y})\,\vert 
        \delta\Psi^{q\bar{q}}_{\rm inel}({\bf x},{\bf y})\rangle
        = \nonumber\\
  && \qquad  1-  \langle\langle\frac{1}{N}
    {\rm Tr}\left[ W^{F\dagger}({\bf x})\, 
                   W^{F}({\bf y})\right] \rangle\rangle
   \langle\langle\frac{1}{N} {\rm Tr}\left[ W^{F\dagger}({\bf y})\, 
                   W^{F}({\bf x})\right]\rangle\rangle  \, ,
   \label{a.8a}
\end{eqnarray}
while the projectile biased inelastic probability is
\begin{eqnarray}
  &&\langle\delta\Psi^{q\bar{q}\, proj}_{\rm inel}({\bf x},{\bf y})\,\vert 
        \delta\Psi^{q\bar{q}\, proj}_{\rm inel}({\bf x},{\bf y})\rangle
        = \nonumber\\
  && \qquad  1-  \langle\langle\frac{1}{N^2}
    {\rm Tr}\left[ W^{F\dagger}({\bf x})\, 
                   W^{F}({\bf y})\right]
   {\rm Tr}\left[ W^{F\dagger}({\bf y})\, 
                   W^{F}({\bf x})\right]\rangle\rangle  \, .
   \label{a.8b}
\end{eqnarray}
Finally the diffractive probability is
\begin{eqnarray}
  &&\langle\delta\Psi^{q\bar{q}}_{\rm diff}({\bf x},{\bf y})\,\vert 
        \delta\Psi^{q\bar{q}}_{\rm diff}({\bf x},{\bf y})\rangle
        \nonumber \\
  && \quad = \langle\langle\left( \frac{1}{N}
    {\rm Tr}\left[ W^F({\bf x})\, W^{F\dagger}({\bf y})\right] 
    - 1\right)\, \left( \frac{1}{N}
    {\rm Tr}\left[ W^F({\bf y})\, W^{F\dagger}({\bf x})\right] 
    - 1\right)\rangle\rangle\, .
   \label{a.8c}
\end{eqnarray}
For the same expressions including the $q\bar qg$ component, we find:
\begin{eqnarray}
  &&\langle\delta\Psi^{(q\bar{q}+q\bar qg)}_{\rm tot}({\bf x},{\bf y})\,\vert 
        \delta\Psi^{(q\bar{q}+q\bar qg)}_{\rm tot}({\bf x},{\bf y})\rangle
        = \nonumber\\
  && \qquad \langle\langle2-  \frac{1}{N}
    {\rm Tr}\left[ W^{F\dagger}({\bf x})\, 
                   W^{F}({\bf y})\right] 
    - \frac{1}{N} {\rm Tr}\left[ W^{F\dagger}({\bf y})\, 
                   W^{F}({\bf x})\right]\nonumber\\
&&- \frac{1}{2N}
 \ln(x_0/x)F^2\nonumber\\
&&\times\Bigg \lbrace
 {\rm Tr}\left[ W^{F\dagger}({\bf z})\, 
    W^F({\bf x})\right]  {\rm Tr}\left[ W^{F\dagger}({\bf y})\, 
    W^F({\bf z})\right] - N{\rm Tr}\left[ W^{F\dagger}({\bf y})\, 
    W^F({\bf x})\right] \nonumber\\
&&+{\rm Tr}\left[ W^{F\dagger}({\bf x})\, 
    W^F({\bf z})\right]  {\rm Tr}\left[ W^{F\dagger}({\bf z})\, 
    W^F({\bf y})\right] - N{\rm Tr}\left[ W^{F\dagger}({\bf x})\, 
    W^F({\bf y})\right] \Bigg \rbrace\rangle\rangle\, ,
   \label{a.8d}
\end{eqnarray}
\begin{eqnarray}
  &&\langle\delta\Psi^{(q\bar{q}+q\bar qg)}_{\rm inel}({\bf x},{\bf y})\,\vert 
        \delta\Psi^{(q\bar{q}+q\bar qg)}_{\rm inel}({\bf x},{\bf y})\rangle
        \nonumber \\
  && = 1- \frac{1}{N^2}
   \langle\langle{\rm Tr}\left[ W^{F\dagger}({\bf x})\, 
                   W^{F}({\bf y})\right] \rangle\rangle
  \langle\langle{\rm Tr}\left[ W^{F\dagger}({\bf y})\, 
                   W^{F}({\bf x})\right] \rangle\rangle
                 \nonumber\\
  &&\qquad +\ln(x_0/x) F^2 {1\over 2N^2}
  \langle\langle{\rm Tr}\left[ W^{F\dagger}({\bf y})\, 
                   W^{F}({\bf x})\right] \rangle\rangle
  \nonumber\\
  &&\times
 \langle\langle\Bigg\{N
  {\rm Tr}\left[ W^{F\dagger}({\bf x}) W^{F}({\bf y})\right] 
   -{\rm Tr}\left[ W^{F\dagger}({\bf x}) W^{F}({\bf z})\right] 
  {\rm Tr}\left[ W^{F\dagger}({\bf z}) W^{F}({\bf y})\right] 
  \Bigg\}\rangle\rangle
  \nonumber\\
&&\qquad  + \ln(x_0/x) F^2 {1\over2 N^2}
  \langle\langle{\rm Tr}\left[ W^{F\dagger}({\bf x})\, 
                   W^{F}({\bf y})\right] \rangle\rangle
  \nonumber \\
&& \times \langle\langle\Bigg\{N
  {\rm Tr}\left[ W^{F\dagger}({\bf y})\, 
                   W^{F}({\bf x})\right] 
  -{\rm Tr}\left[ W^{F\dagger}({\bf z}) W^{F}({\bf x})\right] 
  {\rm Tr}\left[ W^{F\dagger}({\bf y}) W^{F}({\bf z})\right] \Bigg\}
\rangle\rangle\, ,
\label{a.8e}
\end{eqnarray}
\begin{eqnarray}
  &&\langle\delta\Psi^{(q\bar{q}+q\bar qg)}_{\rm diff}({\bf x},{\bf y})\,\vert 
        \delta\Psi^{(q\bar{q}+q\bar qg)}_{\rm diff}({\bf x},{\bf y})\rangle
  \nonumber \\
 && = \langle\langle 
   \left( \frac{1}{N}
    {\rm Tr}\left[ W^F({\bf x})\, W^{F\dagger}({\bf y})\right] 
    - 1\right)
   \left( \frac{1}{N}
    {\rm Tr}\left[ W^F({\bf y})\, W^{F\dagger}({\bf x})\right] 
    - 1\right)
    \nonumber \\
   && \qquad +  C_F\,\ln(x_0/x) F^2\,
    \Bigg \lbrace -\left( \frac{1}{N}
    {\rm Tr}\left[ W^F({\bf x})\, W^{F\dagger}({\bf y})\right] 
    - 1\right)
  \nonumber \\
  && \qquad \qquad \qquad \qquad \qquad 
   \left( \frac{1}{N}
    {\rm Tr}\left[ W^F({\bf y})\, W^{F\dagger}({\bf x})\right] 
    - 1\right)
    \nonumber\\
&&\qquad \qquad \qquad \qquad +\left( \frac{1}{N^2-1}
     {\rm Tr}\left[ W^{F\dagger}({\bf y}) W^F({\bf z})\right]\, 
     {\rm Tr}\left[ W^{F\dagger}({\bf z}) W^F({\bf x})\right]
  \right. \nonumber \\
  && \qquad \qquad \qquad \qquad \qquad \left. 
      -\frac{1}{N(N^2-1)}
    {\rm Tr}\left[ W^F({\bf x})\, W^{F\dagger}({\bf y})\right] 
    - 1\right)
   \nonumber \\
    &&\qquad \qquad \qquad \qquad \quad \times \left( \frac{1}{N^2-1}
     {\rm Tr}\left[ W^{F\dagger}({\bf x}) W^F({\bf z})\right]\, 
     {\rm Tr}\left[ W^{F\dagger}({\bf z}) W^F({\bf y})\right]
  \right. \nonumber \\
  && \qquad  \qquad \qquad \qquad \qquad \left. 
      -\frac{1}{N(N^2-1)}
    {\rm Tr}\left[ W^F({\bf y})\, W^{F\dagger}({\bf x})\right] 
    - 1\right)\Bigg \rbrace\rangle\rangle\, .
\label{a.8f}
\end{eqnarray}
The virtual photon wave function in the approximation (\ref{4.3}) 
contains at most one quark, one antiquark and one gluon. Thus the 
various differential cross sections are obtained form the appropriate 
number of particles by integration over the impact parameter. For example
\begin{eqnarray}
 \frac{d\sigma_{\rm c}^{q\bar{q}}}{d{\bf p}_1\,  d{\bf p}_2} &=&
  \int d^2{\bf x}\, d^2{\bf y}
   \langle
  \delta\Psi^{q\bar{q}}_{\rm c}({\bf x},{\bf y})\, \vert
  a^\dagger_q({\bf p}_1)a_q({\bf p}_1)a^\dagger_{\bar q}({\bf p}_2)
  a_{\bar q}{\bf p}_2)\vert
        \delta\Psi^{q\bar q}_{\rm c}({\bf x},{\bf y}) 
  \rangle\nonumber\\
&=& \int dzd^2{\bf x}\, d^2{\bf y}\, d^2\bar{\bf x}\, d^2\bar{\bf y}\,
       e^{i{\bf p}_1\,({\bf x}-\bar{\bf x})
          +i{\bf p}_2\,({\bf y}-\bar{\bf y})}
        \nonumber \\
&& \times \psi({\bf \bar x- \bar y},z)\psi^*({\bf x- y},z)
   P_c({\bf \bar x,\bar y, x,y})\, ,
\label{a.1}
\end{eqnarray}
and
\begin{eqnarray}
&&  \frac{d\sigma_{\rm c}^{q\bar{q}g}}{d{\bf p}_1\,  d{\bf p}_2\,
  d{\bf k}} =
  \int d{\bf x}\, d{\bf y}\, d{\bf z}\, 
       d\bar{\bf x}\, d\bar{\bf y}\, d\bar{\bf z}\,
       e^{i{\bf p}_1\,({\bf x}-\bar{\bf x})
          +i{\bf p}_2\,({\bf y}-\bar{\bf y})
          +i{\bf k}\,({\bf z}-\bar{\bf z})}
        \nonumber \\
        && \qquad \qquad\qquad\times  \langle
        \delta\Psi^{q\bar{q}g}_{\rm c}
        ({\bf x},{\bf y},{\bf z})\,
    \vert  a^\dagger_q({\bf p}_1)a_q({\bf p}_1)a^\dagger_{\bar q}({\bf p}_2)
    \nonumber \\
    && \qquad \qquad\qquad\qquad\qquad\times 
    a_{\bar q}{\bf p}_2)a^\dagger({\bf k})a({\bf k})
    \vert  \delta\Psi^{q\bar{q}g}_{\rm c}({\bf x},{\bf y},{\bf z})
        \rangle\nonumber\\
&&=  \int dzd^2{\bf x}\, d^2{\bf y}\, d^2{\bf z}\, 
       d^2\bar{\bf x}\, d^2\bar{\bf y}\, d^2\bar{\bf z}\,
       e^{i{\bf p}_1\,({\bf x}-\bar{\bf x})
          +i{\bf p}_2\,({\bf y}-\bar{\bf y})
          +i{\bf k}\,({\bf z}-\bar{\bf z})} \nonumber\\
&&\qquad \times\psi({\bf \bar x- \bar y},{\bf \bar z},z)
        \psi^*({\bf x-  y},{\bf  z},z)
        P_c({\bf \bar x,\bar y,\bar z, x,y,z})\, ,
   \label{a.2}
\end{eqnarray}
where $a_q$, $a_{\bar q}$ and $a$ are annihilation operators of the quark, 
antiquark and gluon respectively. The wave function 
$ \psi({\bf x- y},z)$ is defined in (\ref{gammastar}) and
$\psi({\bf x- y},{\bf z},z)$ can be read off (\ref{gammastar})
and (\ref{4.1}).
$P_c$ are amplitudes given in the following
in terms of the target gluon field averages.
We give the expressions for
the quark-antiquark-gluon probability 
$P_c({\bf \bar x,\bar y,\bar z, x,y,z})$, 
from which one can obtain
the quark-gluon probability as 
\begin{equation}
P_c({\bf \bar x,\bar y, x,y})=\int d^2{\bf z}
P_c({\bf \bar x,\bar y,z, x,y,z})\, .
\end{equation}
We find
\begin{eqnarray}
   && P_{\rm tot}({\bf \bar x,\bar y,\bar z, x,y,z})
   \nonumber \\
   && \quad = \langle\langle{\cal N}\, \bar{\cal N}\, \left( \frac{1}{N}
    {\rm Tr}\left[ W^{F\dagger}(\bar{\bf x})\, W^{F}(\bar{\bf y})\,
                   W^{F\dagger}({\bf y})\, W^{F}({\bf x})
    \right]     - \frac{1}{N}
    {\rm Tr}\left[ W^{F\dagger}(\bar{\bf x})\, 
                   W^{F}(\bar{\bf y})\right]  \right.
    \nonumber \\
  && \qquad \qquad \left.
    - \frac{1}{N} 
    {\rm Tr}\left[ W^{F\dagger}({\bf y})\, 
                   W^{F}({\bf x})\right] + 1 \right) 
    \nonumber \\
  && \qquad\, + \frac{1}{N}\, F\, \bar{F}\, \Bigg [\, W_{ab}^A({\bf z})\, 
               W_{\bar{a}b}^A(\bar{\bf z})\,  
               {\rm Tr}\left[  W^{F\dagger}(\bar{\bf x})\, T^{\bar{a}}\, 
                               W^{F}(\bar{\bf y})\,  W^{F\dagger}({\bf y})\,
                               T^a\, W^F({\bf x}) 
               \right]
    \nonumber \\
    && \qquad \qquad \qquad - W_{ab}^A({\bf z})\, 
                {\rm Tr}\left[T^{a}\, W^{F}({\bf x})\, T^b\, 
                              W^{F\dagger}({\bf y})\right]
               \nonumber \\
    && \qquad \qquad\qquad 
              - W_{ab}^A(\bar{\bf z})\, 
                {\rm Tr}\left[T^{a}\, W^{F}(\bar{\bf y})\, T^b\, 
                              W^{F\dagger}(\bar{\bf x})\right]
              + \frac{N^2-1}{2} \Bigg ]\,\rangle\rangle \, ,
   \label{a.11}
\end{eqnarray}
\begin{eqnarray}
   && P_{\rm inel}({\bf \bar x,\bar y,\bar z, x,y,z})
               = \langle\langle\frac{{\cal N}\, \bar{\cal N}}{N}\, 
        {\rm Tr}\left[W^{F\dagger}(\bar{\bf x})\,
          W^F(\bar{\bf y})\, W^{F\dagger}({\bf y})\, W^F({\bf x})\right]
        \nonumber \\
        &&\qquad - \left(1+\frac{C_F}{2}\, (F-\bar{F})^2\right)\,
         \frac{{\cal N}^2\, \bar{\cal N}^2}{N}\, 
         {\rm Tr}\left[W^{F\dagger}(\bar{\bf x})\,
          W^F(\bar{\bf y})\right]\, 
          \frac{1}{N}\, {\rm Tr}\left[W^{F\dagger}({\bf y})\,
          W^F({\bf x})\right]
        \nonumber \\
        &&\qquad  + \frac{1}{N} \bar{F}F
        {\rm Tr}\left[W^{F\dagger}(\bar{\bf x})\, T^a\, W^F(\bar{\bf y})\,
          W^{F\dagger}({\bf y})\, T^b\, W^F({\bf x})\right]\,
        W_{ac}^A(\bar{\bf z})\, W_{bc}^A({\bf z}) 
        \nonumber \\
        &&\qquad - \frac{1}{N^2} F^2
        {\rm Tr}\left[W^{F\dagger}(\bar{\bf x})\, W^F(\bar{\bf y})\right]\, 
        {\rm Tr}\left[T^a\, W^{F}({\bf x})\, T^b\, W^{F\dagger}({\bf y})
          \right]\, W_{ab}^A({\bf z})
        \nonumber \\
        &&\qquad - \frac{1}{N^2} \bar{F}^2\,
        {\rm Tr}\left[W^{F}(\bar{\bf y})\, T^b\, W^{F\dagger}(\bar{\bf x})\, 
          T^a \right]\, W_{ab}^A(\bar{\bf z})
        {\rm Tr}\left[W^{F\dagger}({\bf y})\, W^F({\bf x})\right]
\rangle\rangle\, ,
   \label{a.9}
\end{eqnarray}
\begin{eqnarray}
   && P_{\rm diff}({\bf \bar x,\bar y,z, x,y,z})   \nonumber \\
   && \quad = \langle\langle{\cal N}\, \bar{\cal N}\, \left( \frac{1}{N}
    {\rm Tr}\left[ W^F({\bf x})\, W^{F\dagger}({\bf y})\right] 
    - 1\right)\, \left( \frac{1}{N}
    {\rm Tr}\left[ W^F(\bar{\bf y})\, W^{F\dagger}(\bar{\bf x})\right] 
    - 1\right)
    \nonumber \\
   && \qquad +  C_F\, F\, \bar{F}\, 
   \left( \frac{2}{N^2-1}
     {\rm Tr}\left[ W^{F\dagger}({\bf y})\, T^a\, W^F({\bf x})\, T^b\right]\, 
     W_{ab}^A({\bf z}) - 1 \right)
   \nonumber \\
   && \qquad \qquad 
   \times \left( \frac{2}{N^2-1}
     {\rm Tr}\left[ W^{F\dagger}(\bar{\bf x})\, T^{\bar{a}}\, 
                    W^F(\bar{\bf y})\, T^{\bar{b}}\right]\, 
     W_{\bar{a}\bar{b}}^A(\bar{\bf z})\, - 1\right)\rangle\rangle\, .
   \label{a.10}
\end{eqnarray}
Here we use the notations of (\ref{4.2},\ref{4.3}) and 
$\bar F$, $\bar {\cal N}$ denote the analogous expressions at
transverse coordinates $\bar {\bf x}$, $\bar {\bf y}$. All terms 
quadratic in $F$ (or $\bar F$) are understood to contain   
the factor $\ln (x_0/x)$.
Also we have not distinguished between the true and the projectile biased
inelastic amplitude, since it does not matter as long as we are interested in 
particles at rapidity far from the target.

\section{Application of Fierz identities}
\label{appb}

In order to simplify the colour algebra in products of Wilson lines,
we use the relation between adjoint and fundamental Wilson lines,
\begin{equation}
  W_{a\, b}^A({\bf x}) = 2\, {\rm Tr}\left[ T^a\, W^F({\bf x})\, 
  T^b\, W^{F\, \dagger}({\bf x})\right]\, ,
  \label{b.1}
\end{equation}
where the generators of the fundamental representation satisfy
\begin{equation}
  T^a\, T^b = \frac{1}{2\, N}\, \delta_{ab} +
  \frac{1}{2}\, d_{abc}\, T^c + i\,  \frac{1}{2}\, f_{abc}\, T^c\, .
  \label{b.2}
\end{equation}
We make extensive use of
the Fierz identities
\begin{eqnarray}
  T_{i\, j}^a\, T_{k\, l}^a = \frac{1}{2} \left( \delta_{il}\, 
  \delta_{jk} - \frac{1}{N} \delta_{ij}\, \delta_{kl} \right)\, ,
  \label{b.3}\\
   W_{i\, j}^F({\bf x})\, W_{k\, l}^{F\, \dagger}({\bf x}) 
   = \frac{1}{N} \delta_{il}\, 
  \delta_{jk} + 2\, W_{a\, b}^A({\bf x})\, 
   T^a_{i\, l}\, T^b_{k\, j}\, .
   \label{b.4}
\end{eqnarray}
The fundamental Wilson lines in (\ref{b.4}) depend on
the same transverse coordinate. This implies that transverse momentum 
integrations in cross sections of the type (\ref{a.1}) or (\ref{a.2}) 
are needed for this Fierz identity to apply. The simplifications
which we shall obtain in using (\ref{b.4}) are technically identical 
to those obtained in a Feynman diagrammatic language via Mueller's 
technique of cancellation between real and virtual (contact) scattering 
contributions. Indeed, in the Feynman diagrammatic language, the only 
prerequisite for application of Mueller's cancellations is again that 
parton lines are closed at the cut, i.e. that transverse coordinates of 
fundamental Wilson lines are equal in amplitude and complex conjugate 
amplitude~\cite{Wdipole}. The use of (\ref{b.4}) will allow
us to circumvent explicit diagrammatic calculations. 

Using the Fierz identities (\ref{b.4}), one can simplify e.g. 
the following contributions to different DIS observables given
in Appendix~\ref{appa} 
\begin{eqnarray}
  &&{\rm Tr}\left[ W^F({\bf z})\, W^{F\dagger}({\bf y})\right]\,  
    {\rm Tr}\left[ W^F({\bf x})\, W^{F\dagger}({\bf z})\right] 
    \nonumber \\
 && = 
    \frac{1}{N}
    {\rm Tr}\left[ W^F({\bf x})\, W^{F\dagger}({\bf y})\right] 
    + 2\, W_{a\,b}^A({\bf z})\, 
      {\rm Tr}\left[ W^{F\dagger}({\bf y})\, T^a\, 
                     W^{F}({\bf x})\, T^b \right]\, ,
   \label{b.5}
\end{eqnarray}
\begin{eqnarray}
  &&{\rm Tr}\left[ W^F({\bf x})\, W^{F\dagger}({\bf y})\right]\,  
    {\rm Tr}\left[ W^F({\bf y})\, W^{F\dagger}({\bf x})\right] 
    \nonumber \\
 && = 
    1 + W_{a\,b}^A({\bf x})\, W_{a\,b}^A({\bf y})\, ,
   \label{b.6}
\end{eqnarray}
\begin{eqnarray}
        && {\rm Tr}\left[W^{F\dagger}({\bf x})\, W^F({\bf y})\right]\, 
        {\rm Tr}\left[T^a\, W^{F}({\bf x})\, T^b\, W^{F\dagger}({\bf y})
          \right]\, W_{ab}^A({\bf z})
          \nonumber \\
        && = \frac{1}{2\, N}\, W_{ab}^A({\bf z})\, W_{ab}^A({\bf x})\, 
        + \frac{1}{2\, N}\, W_{ab}^A({\bf z})\, W_{ab}^A({\bf y})\, 
        \nonumber \\
  && \qquad + \frac{1}{4}\, W_{ab}^A({\bf z})\, 
    W_{cd}^A({\bf y})\, 
    W_{ef}^A({\bf x})\, 
    \left( d_{ace} - if_{ace}\right)\,
    \left(d_{bdf}+if_{bdf}\right)\, ,
          \label{b.7}
\end{eqnarray}
\begin{eqnarray}
 &&{\rm Tr}\left[ W^{F\dagger}({\bf y})\, T^a\, W^F({\bf x})\, T^b\right]\, 
     W_{ab}^A({\bf z})\,
    {\rm Tr}\left[ W^{F\dagger}({\bf x})\, T^{\bar{a}}\, 
                    W^F({\bf y})\, T^{\bar{b}}\right]\, 
     W_{\bar{a}\bar{b}}^A(\bar{\bf z})
     \nonumber \\
 && = \frac{1}{4\, N^2}\, W_{ab}^A({\bf z})\, W_{ab}^A(\bar{\bf z})
 \nonumber \\
 && \qquad + \frac{1}{8\, N}\, W_{ab}^A({\bf z})\, 
    W_{cd}^A(\bar{\bf z})\, 
    W_{ef}^A({\bf y})\, 
    \left( d_{ace} + if_{ace}\right)\,
    \left(d_{dbf}+if_{dbf}\right)
 \nonumber \\
 &&\qquad  + \frac{1}{8\, N}\, W_{ab}^A({\bf z})\, 
    W_{cd}^A(\bar{\bf z})\, 
    W_{ef}^A({\bf x})\, 
    \left( d_{aec} + if_{aec}\right)\,
    \left(d_{dfb}+if_{dfb}\right)
  \nonumber \\
 && \qquad + \frac{1}{16}\, W_{ab}^A({\bf z})\, 
    W_{cd}^A(\bar{\bf z})\, 
    W_{ef}^A({\bf x})\, W_{gh}^A({\bf y})
  \nonumber \\
 && \qquad  \qquad \times  \Bigg [ \frac{2}{N}\, \delta_{ae}\, \delta_{cg}
            + \left( d_{aem}+if_{aem}\right)\, \left(d_{cgm}+if_{cgm}\right)
            \Bigg ]
  \nonumber \\
 && \qquad  \qquad \times   \Bigg [ \frac{2}{N}\, \delta_{fb}\, \delta_{hd}
            + \left( d_{fbn}+if_{fbn}\right)\, \left(d_{hdn}+if_{hdn}\right)
            \Bigg ]\, .
  \label{b.8}
\end{eqnarray}
%
\section{Target averages}
\label{appc}

To perform medium averages $\langle\langle ...\rangle\rangle$ over 
products of fundamental or adjoint Wilson lines, we expand the
Wilson lines locally for each longitudinal position up to second
order in the colour potential. For the last position $\xi$ of the
Wilson lines, one expands e.g.
\begin{eqnarray}
  W_{ab}^A({\bf z}) &=& V_{ab_1}^A({\bf z})\, 
      \left(\delta_{b_1b} - A^g(\xi,{\bf z})\, f^{b_1gb}
            - \frac{C_A}{2}\, \delta_{b_1b}\, B(\xi,{\bf 0})\right)\, ,
  \label{c.1}\\  
  W^F({\bf z}) &=& V^F({\bf z})\, 
      \left(1 + i\,A^g(\xi,{\bf z})\, T^g
            - \frac{C_F}{2}\, B(\xi,{\bf 0})\right)\, .
  \label{c.2}
\end{eqnarray}
Here, $V^A$ and $V^F$ denote the Wilson lines up to the last position
$\xi$. Also, we have anticipated that in the medium averages
coulour fields are expanded only
up to second order locally in the longitudinal direction,
\begin{eqnarray}
  \langle\langle A^a(\xi,{\bf z})\, A^b(\xi,{\bf z'})\, 
    \rangle\rangle = \delta^{ab}\, B(\xi,{\bf z}-{\bf z'})\, .
  \label{c.3}
\end{eqnarray}
In this way, one obtains e.g. for the average of two
fundamental or adjoint Wilson lines
\begin{eqnarray}
   && \langle\langle\, \frac{1}{N}
    {\rm Tr}\left[ W^{F\dagger}({\bf x})\, 
                   W^{F}({\bf y})\right] 
     \rangle\rangle
   \nonumber \\
   &=& 
   \qquad \qquad \langle\langle\, \frac{1}{N}
    {\rm Tr}\left[ V^{F\dagger}({\bf x})\, 
                   V^{F}({\bf y})\right] 
     \rangle\rangle\, \left(1 -C_F\, \left[  B(\xi,{\bf 0})
         - B(\xi,{\bf x}-{\bf y}) \right] \right)
     \nonumber \\
   &=& \qquad \qquad \exp\left[-C_F\, v({\bf x}-{\bf y})\right]\, ,
   \label{c.4}\\
  && \frac{1}{N^2 - 1}\, 
  \langle\langle
  {\rm Tr}\left[ W^{A\, \dagger}({\bf y})\, W^A({\bf x}) \right]
  \rangle\rangle_t 
  = \exp\left[-C_A\, v({\bf x}-{\bf y})\right]\, .
  \label{c.5}
\end{eqnarray}
Here, the last line is obtained by reexponentiation with the help of
\begin{equation}
  v({\bf x}) = 
  \int\, d\xi\, \left[B(\xi,{\bf 0}) - B(\xi,{\bf x})\right]\, .
  \label{c.6}
\end{equation}
More complicated averages can be obtained in the same way:
\begin{eqnarray}
  &&\langle\langle\, W_{a\,b}^A({\bf x})\, 
      {\rm Tr}\left[ W^{F\dagger}({\bf y})\, T^a\, 
                     W^{F}(\bar{\bf y})\, T^b \right]\, \rangle\rangle
  \nonumber \\
  && = \langle\langle\, V_{a\,b}^A({\bf x})\, 
      {\rm Tr}\left[ V^{F\dagger}({\bf y})\, T^a\, 
                     V^{F}(\bar{\bf y})\, T^b \right]\, \rangle\rangle
    \nonumber \\
    && \qquad \times \Bigg [ 
         - \frac{C_A}{2} \left[ B(\xi,{\bf 0}) +  B(\xi,{\bf y-\bar{y}})
             -  B(\xi,{\bf \bar{y} - x}) -  B(\xi,{\bf {y} - x})\right]
     \nonumber \\
    && \qquad \qquad 
    - C_F \left[ B(\xi,{\bf 0}) - B(\xi,{\bf y-\bar{y}})\right]
    \Bigg ]
    \nonumber \\
   && = \frac{N^2-1}{2} e^{-\frac{C_A}{2}\left(
          v({\bf x}-\bar{\bf y}) + v({\bf x}-{\bf y}) \right)
        - \left(C_F - \frac{C_A}{2}\right)\, v({\bf y}-\bar{\bf y})} \, .
      \label{c.7}
\end{eqnarray}
As soon as colour octets propagate along the products of Wilson lines,
the calculations can become more involved. For the combinations of
$f$-symbols
\begin{equation}
  {\cal F}_{b_1d_1}^{bd} \equiv f_{b_1nb}\, f_{d_1nd}\, ,
  \label{c.8}
\end{equation}
which appear in the averaged Wilson lines, it is useful to observe
that both singlet and octet structures are eigenstates of propagation,
\begin{eqnarray}
  {\cal F}_{b_1d_1}^{bd}\, \delta_{bd} &=& C_A\, \delta_{b_1d_1}\, ,
  \label{c.9}\\
  {\cal F}_{b_1d_1}^{bd}\, f_{bdf} &=& \frac{C_A}{2}\, f_{b_1d_1f}\, ,
  \label{c.10}\\
  {\cal F}_{b_1d_1}^{bd}\, d_{bdf} &=& \frac{C_A}{2}\, d_{b_1d_1f}\, .
  \label{c.11}
\end{eqnarray}
For an average of three Wilson lines, this leads to
\begin{eqnarray} 
    &&\left(d_{ace} - if_{ace}\right)\, 
      \langle\langle\, W_{ab}^A({\bf z})\, 
    W_{cd}^A(\bar{\bf z})\, 
    W_{ef}^A({\bf y}) \rangle\rangle 
    \, \left(d_{bdf} - if_{bdf}\right)
    \nonumber \\
    && = \left(d_{ace} - if_{ace}\right)\, 
    \langle\langle\, V_{ab_1}^A({\bf z})\, 
    V_{cd_1}^A(\bar{\bf z})\, 
    V_{ef_1}^A({\bf y}) \rangle\rangle 
   \nonumber \\
  && \qquad \times  \Bigg [ 
    - \frac{3}{2}\, C_A\, B(\xi,{\bf 0})\, \delta_{b_1b}
       \, \delta_{d_1d}\, \delta_{f_1f}
       + B(\xi,{\bf z-\bar{z}})\, {\cal F}_{b_1d_1}^{bd}\, \delta_{f_1f}
   \nonumber \\
   &&\qquad \qquad 
       + B(\xi,{\bf z-y})\, {\cal F}_{b_1f_1}^{bf}\, \delta_{d_1d}
       + B(\xi,{\bf \bar{z}-y})\, {\cal F}_{d_1f_1}^{df}\, \delta_{b_1b}
            \Bigg ]\, \left(d_{bdf} - if_{bdf}\right) \nonumber \\
   && =\left(d_{ace} - if_{ace}\right)\,\langle\langle\, V_{ab_1}^A({\bf z})\, 
    V_{cd_1}^A(\bar{\bf z})\, 
    V_{ef_1}^A({\bf y}) \rangle\rangle\, 
    \left(d_{b_1d_1f_1} - if_{b_1d_1f_1}\right)
   \nonumber \\
  && \times  \Bigg [ 
    - \frac{3\, C_A}{2}\, B(\xi,{\bf 0})
       + \frac{C_A}{2}\, B(\xi,{\bf z-\bar{z}})\, 
       + \frac{C_A}{2}\, B(\xi,{\bf z-y})\, 
       + \frac{C_A}{2}\, B(\xi,{\bf \bar{z}-y})\, 
            \Bigg ]
   \nonumber \\
   && = \left(N^2-1\right)\, \left(2N - \frac{4}{N}\right)\, 
   \exp\left[ -\frac{C_A}{2}\left[ v({\bf z-\bar{z}}) +
              v({\bf z-y}) + v({\bf \bar{z}-y})\right] \right]\, .
  \label{c.12}
\end{eqnarray}
More complicated is the average over the four adjoint Wilson
lines given in (\ref{b.8}). Again, we expand the Wilson lines
according to (\ref{c.1}),
\begin{eqnarray}
  &&\langle\langle\, W_{ab}^A({\bf z})\, W_{cd}^A({\bf z})\, 
    W_{ef}^A({\bf x})\, W_{gh}^A({\bf y})\rangle\rangle
    \nonumber \\
  &&\quad  
  = \langle\langle\, V_{a\bar{a}}^A({\bf z})\, V_{c\bar{c}}^A({\bf z})\, 
    V_{e\bar{e}}^A({\bf x})\, V_{g\bar{g}}^A({\bf y})\rangle\rangle\,
    M^{\bar{a}\bar{c}\bar{e}\bar{g}}_{bdfh}\, .
    \label{c.13}
\end{eqnarray}
We introduce the shorthands
\begin{eqnarray}
  R_a = - v({\bf z}-{\bf x}) - v({\bf z}-{\bf y})\, ,\qquad
  R_b = - v({\bf x}-{\bf y})\, .
  \label{c.14}
\end{eqnarray}
One eigenstate of $M^{aceg}_{bdfh}$ is rather easily checked,
\begin{equation}
  M^{aceg}_{bdfh}\, \left(f_{fbn}d_{hdn} + d_{fbn}f_{hdn}\right)
  = R_a\, \left(f_{ean}d_{gcn} + d_{ean}f_{gcn}\right)\, .
  \label{c.15}
\end{equation}
To find the other eigenstates of $M^{aceg}_{bdfh}$, we introduce
the vectors
\begin{eqnarray}
  u_1 &=& \delta_{ea}\delta_{gc}\, ,\quad
  u_2 = \delta_{ca}\delta_{ge}\, ,\quad
  u_3 = \delta_{ga}\delta_{ec}\, ,
  \label{c.16} \\
  v_1 &=& f_{ean}f_{gcn}\, ,\quad
  v_2 = f_{can}f_{gen}\, ,\quad
  v_3 = f_{gan}f_{ecn}\, ,
  \label{c.17} \\
  w_1 &=& d_{ean}d_{gcn}\, ,\quad
  w_2 = d_{can}d_{gen}\, ,\quad
  w_3 = d_{gan}d_{ecn}\, .
  \label{c.18}
\end{eqnarray}
Colour algebra implies the identities
\begin{eqnarray}
  v_1 &=& \frac{2}{N}\left( u_2-u_3\right) + w_2 - w_3\, ,
  \label{c.19} \\
  v_2 &=& \frac{2}{N}\left( u_1-u_3\right) + w_1 - w_3\, ,
  \label{c.20} \\
  v_3 &=& \frac{2}{N}\left( u_2-u_1\right) + w_2 - w_1\, .
  \label{c.21}
\end{eqnarray}
In the six-dimensional space, given by the vectors
$\left(u_1,u_2,u_3,w_1,w_2,w_3\right)$, one finds by explicit 
calculation that the matrix $M^{aceg}_{bdfh}$ acts like
(using $R_d = R_b-R_a$)
\begin{equation}
  \pmatrix{ NR_a & 0 & -\frac{2}{N}R_d & 0 & 0 & 
            R_d\, \left(\frac{4}{N^2} - 1\right)\cr
            \frac{2}{N}R_d & NR_b & \frac{2}{N}R_d & 
            R_d\, \left(\frac{4}{N^2} - 1\right) & 0 &
            R_d\, \left(\frac{4}{N^2} - 1\right)\cr
            -\frac{2}{N}R_d & 0 & NR_a & R_d\, \left(\frac{4}{N^2} - 1\right)
            & 0 & 0\cr
            0 & 0 & R_d & \frac{N}{4}\left(3R_a+R_b\right) & 0 &
            R_d\left(\frac{2}{N} - \frac{N}{4}\right)\cr
            R_d & 0 & R_d & -R_d\left(\frac{2}{N} - \frac{N}{4}\right) &
            \frac{N}{2}\left(R_a+R_b\right) & 
            -R_d\left(\frac{2}{N} - \frac{N}{4}\right)\cr 
            R_d & 0 & 0 & R_d\left(\frac{2}{N} - \frac{N}{4}\right) &
            0 & \frac{N}{4}\left(3R_a+R_b\right)\cr}
  \label{c.22}
\end{equation}
To calculate the average in (\ref{b.8}), we have to
propagate the vector
\begin{equation}
  \Bigg [ \frac{2}{N}\, \delta_{fb}\, \delta_{hd}
            + d_{fbn}d_{hdn} - f_{fbn}f_{hdn} \Bigg ]
         = \frac{2}{N}\left( u_1 - u_2 + u_3\right)
         + w_1 - w_2 + w_3\, .
  \label{c.23}
\end{equation}
Finding the eigenvectors and eigenvalues of (\ref{c.22}) with
the help of Mathematica, decomposing the vector (\ref{c.23})
in terms of the eigenvectors of (\ref{c.22}) and reexponentiating
the obtained expressions, we find
\begin{eqnarray}
 && \frac{1}{16}\, \langle\langle W_{ab}^A({\bf z})\, 
    W_{cd}^A({\bf z})\, 
    W_{ef}^A({\bf x})\, W_{gh}^A({\bf y})\rangle\rangle
  \nonumber \\
 && \qquad  \qquad \times  \Bigg [ \frac{2}{N}\, \delta_{ae}\, \delta_{cg}
            + \left( d_{aem}+if_{aem}\right)\, \left(d_{cgm}+if_{cgm}\right)
            \Bigg ]
  \nonumber \\
 && \qquad  \qquad \times   \Bigg [ \frac{2}{N}\, \delta_{fb}\, \delta_{hd}
            + \left( d_{fbn}+if_{fbn}\right)\, \left(d_{hdn}+if_{hdn}\right)
            \Bigg ]
   \nonumber \\
 &=& \frac{1}{4N^2} e^{NR_b} + \frac{(N-1)(N+3)N^2}{16} e^{NR_a + R_a - R_b}
     + \frac{N^2-1}{N^2} e^{\frac{N}{2}(R_a+R_b)} 
     \nonumber \\
     &&      + \frac{(N+1)(N-3)N^2}{16}e^{NR_a - R_a + R_b} +
     \frac{N(N^2-1)\left(N-\frac{4}{N}\right)}{8}\, e^{NR_a}\, .
 \label{c.24}
\end{eqnarray}
This enters the calculation of the probability 
$P_{\rm diff}^{(q\bar{q} + q\bar{q}g)}$ in (\ref{4.14a}) and the
diffractive evolution equation (\ref{5.8}).

Finally we present averaged expressions for the single differential quark 
cross sections, obtained from (\ref{a.11},\ref{a.10}) by taking $\bar y=y$.
\begin{equation}
  \frac{d\sigma_{\rm tot}^{q\bar{q}}}{d{\bf p}} =
  \int d{\bf x}\,  d{\bf y}\, d\bar{\bf x}\,  
       e^{i{\bf p}\,({\bf x}-\bar{\bf x})}
  \left( 1+ 
    e^{-C_F\, v(\bar{\bf x}-{\bf x})}-
    e^{-C_F\, v(\bar{\bf x}-{\bf y})}-
    e^{-C_F\, v({\bf x}-{\bf y})}   \right)\, ,
\label{d.2}
\end{equation}
\begin{eqnarray}
  \frac{d\sigma_{\rm diff}^{q\bar{q}}}{d{\bf p}} 
   &=&  \int d{\bf x}\, d{\bf y}\, d\bar{\bf y}\, 
       e^{i{\bf p}\,({\bf y}-\bar{\bf y})}
       \left( \frac{1}{N^2} e^{-C_F\, v(\bar{\bf y}-{\bf y})}
         - e^{-C_F\, v(\bar{\bf x}-{\bf y})} 
         - e^{-C_F\, v(\bar{\bf x}-\bar{\bf y})} 
         \right.
         \nonumber \\ && \qquad \left.
       +  \frac{N^2-1}{N^2} e^{-\frac{C_A}{2}\left(
          v({\bf x}-\bar{\bf y}) + v({\bf x}-{\bf y}) \right)
        - \left(C_F - \frac{C_A}{2}\right)\, v({\bf y}-\bar{\bf y})} 
        + 1 \right)\, .
   \label{d.1}
\end{eqnarray}
%


\end{document}